\documentclass[aps,prl,reprint,superscriptaddress,amsmath,amssymb,floatfix]{revtex4-2}

\usepackage{graphicx}  
\usepackage{dcolumn}   
\usepackage{bm}        
\usepackage{comment}
\usepackage{booktabs}  

\usepackage{physics}   

\usepackage[colorlinks,hyperindex]{hyperref}

\usepackage{placeins}


\begin{document}

\title{Strong Angular-Momentum-Coupling Dependence \\of Atomic Decay
Angular Distributions in a Hybrid $\mathbf{LS/LSJ}$ Framework}

\author{T.~J.~M. Zouros}
\email{tzouros@physics.uoc.gr}
\affiliation{Department of Physics, University of Crete, GR-70013 Heraklion, Greece}

\date{\today}

\begin{abstract}
State production and its subsequent decay can each require a different
angular-momentum coupling treatment.
Although long recognized, this
distinction is often overlooked in the two-step model, where $LS$
coupling is traditionally applied to both production and decay
($LS/LS$), even when the decay requires $LSJ$ coupling.
We introduce a hybrid $LS/LSJ$ formulation that uses established
angular-momentum recoupling theory to project the initial $LS$ alignment
onto the fine-structure $J$ states, evaluates their decay in $LSJ$
coupling, and, for well-separated $J$ levels, sums their contributions
incoherently.
The $1s2s2p\,{}^{4}\!P_J$ manifold produced by single-electron capture
in $\mathrm{C}^{4+}(1s2s\,{}^{3}\!S)+\mathrm{He}$ collisions provides
a good test case, with $M_L$-resolved theoretical production
cross sections and absolute measurements of the Auger
single-differential cross section available. Using the same $LS$
production cross sections in both treatments, we find that the hybrid
$LS/LSJ$ treatment strongly suppresses and can even
invert the Auger anisotropy relative to the traditional $LS/LS$ treatment,
substantially improving agreement with the measurements. Such an
incoherent hybrid formulation provides the appropriate framework when $LS$
coupling governs the initial interaction dynamics while well-separated
fine-structure levels require an $LSJ$ description of the decay.
\end{abstract}
\maketitle

Angular-momentum polarization, encompassing alignment and orientation,
arises whenever magnetic substates are populated anisotropically, occurring
across widely different areas of physics and chemistry, from charged-particle
collisions to photon-induced processes \cite{fan73a,and17a,zar26a}. It
governs the polarization and angular distributions of emitted photons and
electrons and provides a means of probing and controlling atomic and
molecular dynamics. Modern synchrotron and free-electron-laser sources have
extended such studies to the controlled preparation and probing of aligned
states and to ultrafast polarization-dependent dynamics
\cite{you06a,zhu09a,kan19a,per23a}.

In atomic physics using directed beams, state-selective production of excited
atoms and ions provides important benchmarks for atomic interaction dynamics
and structure theories~\cite{dra06a,fan73a,bec96a}. Ion, electron, and photon
beams can produce non-statistical populations of the magnetic
substates~\cite{per58a,mehl68c,ber78a}, whose alignment is reflected in the
angular distributions of subsequent relaxation products, including Auger
electrons~\cite{cle74a,eic76a,ber77a} and photons~\cite{eic76a,ber77a}.

The angular distribution of these relaxation products depends on the
angular-momentum coupling appropriate to both state production and its subsequent
decay~\cite{bal00b}. When the electrostatic Coulomb interaction dominates, its spin
independence and commutation with the total orbital and spin angular momenta
$L$ and $S$ make $LS$ coupling the natural description. This need not remain
true for the subsequent decay: if the Coulomb decay is forbidden by selection
rules, weaker non-Coulombic interactions that were negligible during production
can instead provide the leading decay mechanism. Production and decay of the
same atomic state can therefore require different angular-momentum coupling
treatments.

The appropriate treatment of the decay also depends on the separation of the
fine-structure levels $\Delta E_\text{FS}$ relative to their natural widths
$\Gamma$. When $\Delta E_\text{FS}\ll\Gamma$, the levels overlap strongly
and their decay amplitudes remain coherent, corresponding to the $LS$ decay
limit~\cite{mehl80a,kab07a}. Conversely, when $\Delta E_\text{FS}\gg\Gamma$, the
fine-structure levels are well separated and form independent decay channels
whose angular distributions are described in $LSJ$ coupling~\cite{cle74a,kab07a}.
For partially overlapping levels, $\Delta E_\text{FS}\sim\Gamma$, Mehlhorn
and Taulbjerg~\cite{mehl80a} (MT) accounted for coherence between the
fine-structure components by introducing a time-integrated de-alignment
parameter $D_k$. The MT treatment provides the connection between the
coherent $LS$ limit and the incoherent fine-structure limit as the degree
of level overlap changes~\cite{cle74a,mehl80a}.

The sensitivity of Auger angular distributions (AADs) to angular-momentum
coupling has been extensively studied for target atoms and
ions~\cite{cle74a,mehl80a,kab94a,ma20a,wu24a}. For excited projectile ions,
zero-degree Auger projectile spectroscopy (ZAPS) provides high-resolution,
state-selective measurements of absolute single differential cross sections
(SDCS)~\cite{sto87a,zou97a}. Selection of the projectile charge state provides
access to few-electron systems amenable to calculations and systematic
isoelectronic investigations.

Among the few-electron systems accessible to ZAPS, the long-lived carbon
$1s2s2p\,^4\!P_J$ multiplet populated via single electron capture
(SEC)~\cite{mad22a} provides a particularly suitable benchmark. The
same $1s2s2p\,^4\!P_J$ multiplet has also been studied following
$1s\rightarrow2p$ excitation of Li-like $1s^22s$
projectiles~\cite{zou89b,lee91a} and K-shell ionization of metastable Be-like
$1s^22s2p\,^3\!P$ projectiles~\cite{lee92a}.

For the carbon SEC benchmark, production is dominated by the electrostatic Coulomb interaction and is therefore appropriately described in $LS$ coupling. The magnetic-substate production cross sections were obtained from nonperturbative three-electron atomic-orbital close-coupling (AOCC) calculations within a full configuration interaction approach~\cite{mad22a}. Coulomb autoionization of the $^4\!P$ term to the $1s^2\,{}^1\!S_0$ ground state is spin forbidden, and the decay is enabled by weaker relativistic interactions~\cite{dav89a}.
The fine-structure components are well isolated
and non-overlapping, $\Delta E_\text{FS}\gg\Gamma$, requiring an $LSJ$
description of their decay, whereas the previous analysis of this system
employed a pure $LS$ description of the angular
distribution~\cite{mad22a}. A state-resolved description raises additional
angular-momentum considerations that are absent in pure $LS$ coupling. For
example, the $J=5/2$ component can in principle introduce an additional
$P_4(\cos\theta)$ term in the Legendre expansion of the AAD that is absent
in the pure $LS$ treatment~\cite{sur08a,fri12a}.

In this work, we present a general hybrid $LS/LSJ$ coupling framework for
systems in which production is appropriately described in $LS$ coupling
while the subsequent decay proceeds through isolated fine-structure states,
$\Delta E_\text{FS}\gg\Gamma$. The alignment produced in $LS$ coupling is
projected onto the fine-structure $J$ levels before their decay angular
distributions are evaluated in $LSJ$ coupling.
Applied to the carbon benchmark of Ref.~\cite{mad22a}, the hybrid framework
substantially modifies the predicted AAD, including suppression and inversion
of the anisotropy, and improves agreement with the absolute measurements
while retaining the same original $LS$ production cross sections.
More generally, our results show the
importance of treating state production and its subsequent decay according to
the angular-momentum coupling scheme appropriate to each step. Detailed
derivations, parameter tables, and additional figures are provided in the
Supplemental Material~\cite{sm26a}.

Production and autoionization are described within the independent two-step
model~\cite{loh09a}, in which state preparation and decay are treated separately
and linked through the statistical description of the intermediate
state~\cite{blu12a}. Here, we restrict the treatment to the isolated-resonance
approximation; overlapping resonances would require a more general formulation.

The standard two-step AAD formalism, in which the angular-distribution
coefficients factorize into the production alignment $\mathcal{A}_{k0}$ and
intrinsic decay anisotropy $\alpha_k$, is given by
Eq.~\eqref{eq:twostep_factorization} in the End Matter, with the
single-partial-wave expressions in Eqs.~\eqref{eq:alpha_J_unreduced}
and \eqref{eq:alphak_singlel}. We now proceed to the $LS/LSJ$ hybrid
construction.

The $LS/LSJ$ hybrid scheme applies to fine-structure multiplets for which
state production is governed by $LS$ coupling, while the subsequent decay
proceeds through individual $J$ fine-structure channels.
Its physical basis
is the separation of production and decay inherent in the two-step model.
For the present $^4\!P_J$ manifold, the isolated-resonance condition
$\Delta E_{\rm FS}\gg\Gamma$ is strongly satisfied, with even the smallest
ratio $|\Delta E_{JJ'}|/\Gamma_{JJ'}=3.44\times10^3$ (see
Table~\ref{tb:hybridcriterion}, End Matter).

The collision-produced alignment $\mathcal{A}_{k0}[L]$ is naturally expressed
in terms of the orbital angular momentum $L$. To describe the subsequent decay
through the individual fine-structure channels, this orbital alignment is
transformed into the coupled $J$ representation using standard
angular-momentum recoupling~\cite{kab94a}:
{\small
\begin{equation}
\mathcal{A}_{k0}^{\rm proj}[L,S,J] =
(-1)^{L+S+J+k}\hat{L}\hat{J}
\begin{Bmatrix}
L & J & S \\
J & L & k
\end{Bmatrix}
\mathcal{A}_{k0}[L].
\label{eq:calAk0_proj_def}
\end{equation}
}%
Equation~\eqref{eq:calAk0_proj_def} represents a purely geometrical
transformation of the collision-produced orbital alignment and introduces no
additional collision dynamics. Alternatively, the corresponding
fine-structure magnetic-sublevel cross sections are obtained directly from
the $LS$ partial cross sections by the Percival--Seaton
transformation~\cite{per58a},
{\small
\begin{equation}
\sigma(J,M_J)=
\frac{\hat{J}^{2}}{\hat{S}^{2}}
\sum_{M_L,M_S}
\begin{pmatrix}
S & L & J\\
M_S & M_L & -M_J
\end{pmatrix}^{2}
\sigma(L,|M_L|).
\label{eq:sigmaJMJ_to_sigmaLML}
\end{equation}
}%
The two transformations are equivalent: substitution of
Eq.~\eqref{eq:sigmaJMJ_to_sigmaLML} into the definition of
$\mathcal{A}_{k0}[J]$ yields Eq.~\eqref{eq:calAk0_proj_def}.
This equivalence, which does not appear to have been noted explicitly,
is demonstrated in the SM.

For each fine-structure channel, the hybrid angular-distribution coefficient
is therefore
\begin{equation}
a_k^\text{hybrid}[L,S,J] =
\alpha_k^{(j=J)}[J,J_f=0]\,
\mathcal{A}_{k0}^{\rm proj}[L,S,J],
\label{eq:ak_hybrid}
\end{equation}
where $\alpha_k^{(j=J)}[J,J_f=0]$ is given by
Eq.~\eqref{eq:alpha_J_unreduced}. We now apply this formulation to the
benchmark $1s2s2p\,^4\!P_J$ manifold and combine the individual $J$-channel
contributions to obtain the experimentally unresolved multiplet AAD.


The $1s2s2p\,^4\!P_J$ state is produced primarily via $2p$ SEC
into the metastable $1s2s\,^3\!S$ component of the projectile
beam~\cite{mad22a,mad20a}, with additional cascade contributions from
higher-lying states~\cite{zou20a}. To compare directly with available ZAPS
measurements, we specialize the AAD to $\theta=0^\circ$. The $^4\!P_J$ levels
are long-lived and have markedly different lifetimes
($\tau_{1/2}=2.94$~ns, $\tau_{3/2}=7.10$~ns, and
$\tau_{5/2}=121.36$~ns; see Table~S6 in the SM).
Consequently, the Auger electrons are emitted in flight as the projectile
ions travel from the target cell toward the spectrometer entrance, producing
an extended rather than point-like emission source. This transforms the
point-source solid angle $d\Omega_0$ into a level-dependent effective solid
angle $d\Omega_J=G_J\,d\Omega_0$. The solid-angle correction factors $G_J$
account for two competing in-flight effects~\cite{zou97a}: the geometric
solid-angle gain as the emitting ions approach the spectrometer entrance and
the exponential beam depletion following the collision. Their calculation and numerical values are given in the SM.

In the traditional pure $LS$-coupling model, the experimentally unresolved $^4\!P$
multiplet is described by the $L=1$ orbital alignment. A $J$-averaged solid-angle correction factor $\overline{G}_\tau$
is conventionally applied~\footnote{We retain the notation
$\overline{G}_\tau$ used in Ref.~\cite{mad22a}; the level-specific factors
$G_{\tau_J}$ are denoted more compactly as $G_J$ here.}.
The prompt-equivalent, zero-degree $LS$ AAD is therefore:
\begin{equation}
\frac{d\sigma^{LS}_A[^4\!P]}
{\overline{G}_\tau\,d\Omega_0}(0^\circ)
=
\overline{\xi}\,\frac{\sigma[^4\!P]}{4\pi}
\Bigl(1+\alpha_2\mathcal{A}_{20}\Bigr),
\label{eq:final_LS_0deg}
\end{equation}
where the Legendre polynomial, $P_2(1)=1$, and
$\alpha_2$ and $\mathcal{A}_{20}$ are defined for $L=1$ by
Eqs.~\eqref{eq:alphak_singlel} and \eqref{eq:calAk0L_def}.

In the hybrid model, the level-specific solid-angle correction factors $G_J$
are applied separately to each fine-structure component. The prompt-equivalent,
zero-degree AAD for each $J$ component is:
{\small
\begin{equation}
\frac{d\sigma^{\text{hybrid}}_A[J]}
{G_J\,d\Omega_0}(0^\circ)
=
\xi_J\,\frac{\sigma[J]}{4\pi}
\left[
1+\sum_{k=2,4}\alpha_k^{(J)}
\mathcal{A}_{k0}^{\rm proj}[J]
\right].
\label{eq:hybrid_AAD_generic_J}
\end{equation}
}%
The channel-resolved expressions are given in
Sec.~A.3 of the SM.
Equation~\eqref{eq:hybrid_AAD_generic_J} shows the distinct anisotropy
structure of the individual fine-structure channels. The $J=1/2$ component
cannot support alignment~\cite{cle74a} and is therefore strictly isotropic
($\mathcal{A}^\text{proj}_{20}[1/2]=\mathcal{A}^\text{proj}_{40}[1/2]=0$), while for
$J=3/2$ the expansion truncates at rank $k=2$
($\alpha_4^{(3/2)}=0$). The $J=5/2$ component formally supports multipoles
up to $k=4$. For the $L=1$ orbital alignment considered here, however,
$\mathcal{A}^\text{proj}_{40}[5/2]=0$, so that the overall multiplet
distribution also truncates at $k=2$.

For the isolated, non-overlapping fine-structure levels considered here, the
unresolved AAD is obtained within the incoherent hybrid $LS/LSJ$ treatment by
summing the individual fine-structure contributions~\cite{kab94a},
{\small
\begin{align}
\frac{d\sigma^{\text{hybrid}}_A[^4\!P]}{d\Omega_0}&(0^\circ)
\equiv
\sum_J
\frac{d\sigma^{\text{hybrid}}_A[J]}{d\Omega_0}(0^\circ),
\label{eq:AAD0_LSJ_hybrid_GtauJ_sumJ_def}\\
&=
\sum_J G_J\,\xi_J\,\frac{\sigma[J]}{4\pi}
\left[
1+\!\!\!\sum_{k=2,4}\alpha_k^{(J)}
\mathcal{A}_{k0}^{\rm proj}[J]
\right].
\end{align}
}%
The corresponding $J$-averaged Auger yield $\overline{\xi}$ and solid-angle
correction factor $\overline{G}_\tau$ are~\cite{zou97a}:
\begin{equation}
\overline{\xi} =
\frac{\sum_J \hat{J}^2 \xi_J}{\sum_J\hat{J}^2},
\quad \text{and} \quad
\overline{G}_\tau =
\frac{\sum_J \hat{J}^2 G_J \xi_J}{\sum_J \hat{J}^2 \xi_J}.
\label{eq:barGLxi}
\end{equation}
The additional $\xi_J$ weighting in $\overline{G}_\tau$ follows from matching
the $J$-summed and
pure-$LS$ expressions. Using the statistical relation between the
fine-structure and configuration production cross sections, the same
$\sigma[^4\!P]$ is retained in both representations:
{\small
\begin{equation}
\overline{G}_\tau\cdot\overline{\xi}\cdot\sigma[^4\!P]
=
\sum_J G_J\xi_J\,\sigma[^4\!P_J], \quad
\sigma[^4\!P_J]=\frac{\hat{J}^2\,\sigma[^4\!P]}{\sum_J\hat{J}^2}.
\label{eq:total_sigma_A_equality}
\end{equation}
}%
Here $\sum_J\hat{J}^2=\hat{L}^2\hat{S}^2=3\cdot4=12$ (see
Sec.~D of the SM).

Substituting Eqs.~\eqref{eq:calAk0_proj_def} and
\eqref{eq:total_sigma_A_equality} into
Eq.~\eqref{eq:AAD0_LSJ_hybrid_GtauJ_sumJ_def} yields the
prompt-equivalent form for the entire multiplet:
\begin{equation}
\frac{d\sigma^{\text{hybrid}}_A[^4\!P]}
{\overline{G}_\tau\,d\Omega_0}(0^\circ)
=
\overline{\xi}\,\frac{\sigma[^4\!P]}{4\pi}
\Bigl(1+\alpha_2\mathcal{A}_{20}\overline{H}_2\Bigr),
\label{eq:final_hybrid_0deg}
\end{equation}
where $\overline{H}_2$ is the $J$-averaged hybrid coupling factor.

For an individual fine-structure channel, the hybrid coupling factor $H_k[L,S,J]$ for an arbitrary
$^{2S+1}L_J$ multiplet is [see
Sec.~A.6 of the SM]:
{\small
\begin{equation}
H_k[L,S,J] =
(-1)^{S+\frac{1}{2}}
\hat{J}^2
\begin{Bmatrix}
L & J & S \\
J & L & k
\end{Bmatrix}
\frac{
\begin{pmatrix}
J & J & k \\
\frac{1}{2} & -\frac{1}{2} & 0
\end{pmatrix}
}{
\begin{pmatrix}
L & L & k \\
0 & 0 & 0
\end{pmatrix}
}.
\label{eq:Hk_universal}
\end{equation}
}%
The $J$-averaged hybrid coupling factor is then defined as
\begin{equation}
\overline{H}_k[^{2S+1}L] \equiv
\frac{\sum_J G_J\,\xi_J\,\hat{J}^2\,H_k[L,S,J]}
{\sum_J G_J\,\xi_J\,\hat{J}^2},
\label{eq:Hk_macroscopic_avg}
\end{equation}
with $\sum_JG_J\xi_J\hat{J}^2
=\overline{G}_\tau\,\overline{\xi}\,\sum_J\hat{J}^2$
from Eq.~\eqref{eq:barGLxi}.

For the $1s2s2p\,^4\!P$ configuration ($L=1,S=3/2$), evaluation of
Eq.~\eqref{eq:Hk_universal} at $k=2$ gives
$H_2[1/2]=0$, $H_2[3/2]=-2/5$, and $H_2[5/2]=2/5$
[see Eqs.~(S65)--(S67) in the SM].
Using these values and $\sum_J\hat{J}^2=12$ in
Eq.~\eqref{eq:Hk_macroscopic_avg} gives
[see Eq.~(S38) in the SM]:
\begin{equation}
\overline{H}_2(E_p) =
\frac{1}{15}\left(
\frac{3G_{5/2}\xi_{5/2}-2G_{3/2}\xi_{3/2}}
{\overline{G}_\tau\,\overline{\xi}}
\right).
\label{eq:H2_overline_def}
\end{equation}
Here, the $E_p$ projectile-energy dependence arises from the solid-angle correction
factors. In the ideal unweighted limit, $G_J\rightarrow1$ and
$\xi_J\rightarrow\overline{\xi}$, giving
$H_2^\text{ideal}=1/15$.

The solid-angle correction factors $G_J$ were recalculated
using the effective solid-angle formalism introduced by Doukas
\textit{et al.}~\cite{dou15a} and subsequently applied in
Ref.~\cite{ben18a}, which combines SIMION~\cite{simion} Monte Carlo
ray-tracing of the spectrometer acceptance with analytical integration.
The multi-configuration Dirac--Fock (MCDF) radiative and Auger rates of
Ref.~\cite{ben15a} were used, from which the lifetimes and Auger yields
were obtained self-consistently, yielding $G_J$ values in close agreement
with those reported by Madesis \textit{et al.}~\cite{mad22a} (see
Sec.~E of the SM). The projectile-energy dependence of the $G_J$ factors
is shown in Fig.~S3.


The prompt-equivalent, zero-degree SDCS can then be written in the unified form:
\begin{equation}
\frac{d\sigma^\text{hybrid}_A}{\overline{G}_\tau \,d\Omega_0}(0^\circ)
=
\overline{\xi}\,\frac{\sigma^\text{tot}}{4\pi}
\Bigl(1+\alpha_2\,\mathcal{A}_{20}^{\text{tot}}\,M_2\Bigr),
\label{eq:AAD_M2_master}
\end{equation}
where $\sigma^\text{tot}$ represents the total
($\text{SEC}+\text{cascades}$) configuration production cross section
from Ref.~\cite{mad22a}, $\overline{\xi}$ is the $J$-averaged Auger
yield, and $\overline{G}_\tau$ is the $J$-averaged solid-angle
correction factor.
The angular-distribution treatments are then distinguished
by $M_2$.

In this representation, the alignment and single-partial-wave
intrinsic anisotropy parameters, $\mathcal{A}_{20}^{\text{tot}}$ and
$\alpha_2=-\sqrt{2}$ [Eq.~\eqref{eq:alphak_singlel}], respectively,
are separated from the fine-structure effects incorporated through
$M_2$. For isotropic cascade feeding~\cite{zou20a}, the total alignment
entering Eq.~\eqref{eq:AAD_M2_master} is obtained by weighting the SEC
alignment by the ratio $\sigma^\text{SEC}/\sigma^\text{tot}$, as derived
in the End Matter, Eq.~\eqref{eq:a2calA20SECtimesratio}. The SEC and
cascade cross sections are those of Madesis
\textit{et al.}~\cite{mad22a}, as listed in
Table~S1 of the SM.

\begin{figure}[htbp]
\centering
\includegraphics[width=\linewidth]{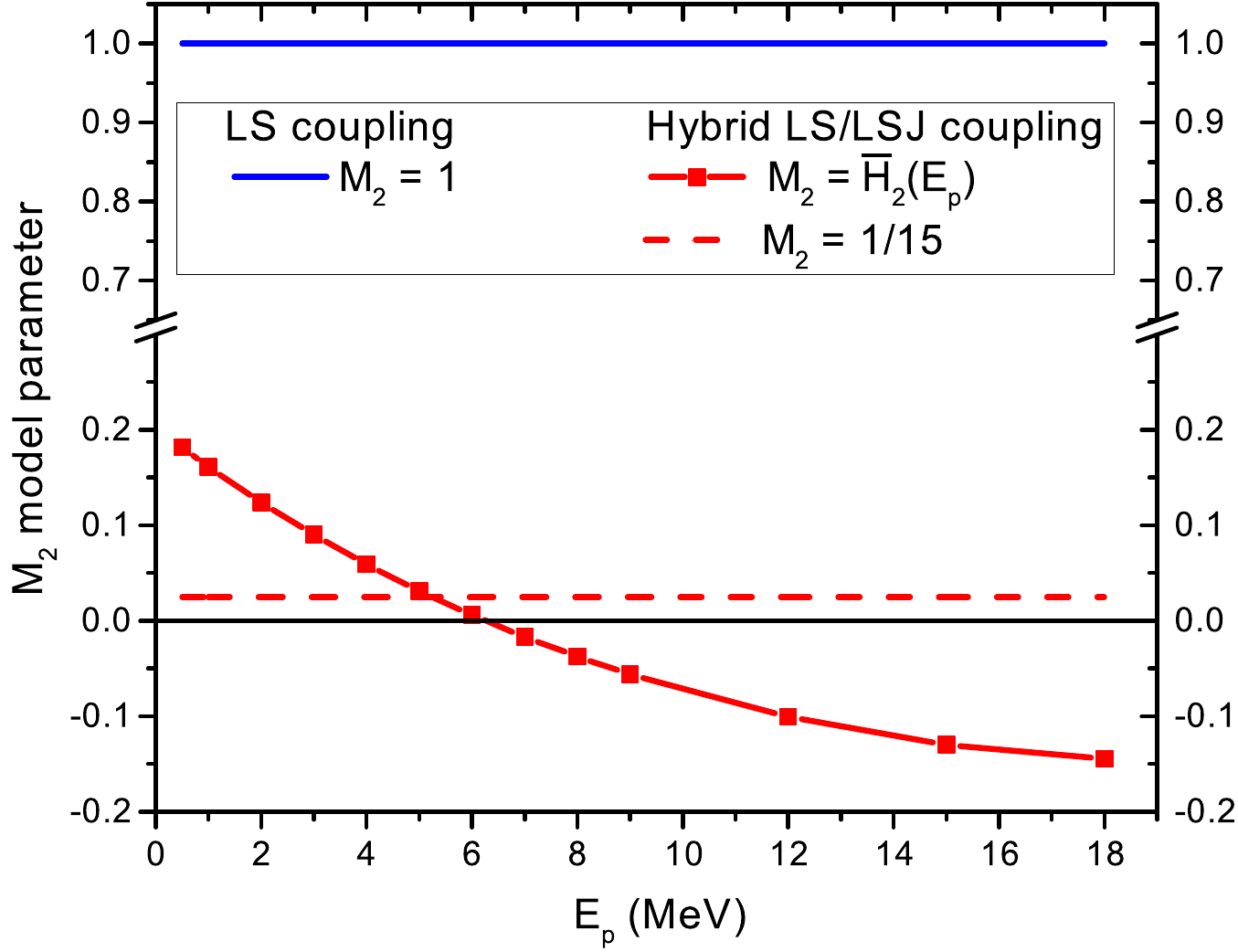}
\caption{The AAD  model parameter $M_2$ versus projectile energy $E_p$
for the unresolved $\text{C}^{3+}(1s2s2p\ ^4\!P)$ multiplet.
The three curves are the $M_2$ values defined in the text.
The hybrid model $\overline{H}_2(E_p)$
[Eq.~\eqref{eq:H2_overline_def}] crosses the $M_2=0$ isotropic
line at $E_p=6.26$~MeV, showing a large disparity with the
traditional pure $LS$ model ($M_2=1$) used in Ref.~\cite{mad22a}.}
\label{fig:alignment_comparison}
\end{figure}

For comparison, we consider the following three values of the model parameter $M_2$ in Eq.~\eqref{eq:AAD_M2_master}:
\begin{enumerate}
\item $M_2 = 1$: the traditional pure $LS$ coupling model, as utilized in Ref.~\cite{mad22a}.
\item $M_2 = \overline{H}_2(E_p)$: the $LS/LSJ$ hybrid model 
Eq.~\eqref{eq:H2_overline_def}.
\item $M_2=H_2^\text{ideal}=1/15$: the ideal unweighted limit of the
present hybrid framework.
\end{enumerate}
The three $M_2$ values are compared as a function of projectile energy
$E_p$ in Fig.~\ref{fig:alignment_comparison}.

\begin{figure}[htbp]
\centering
\includegraphics[width=\columnwidth]{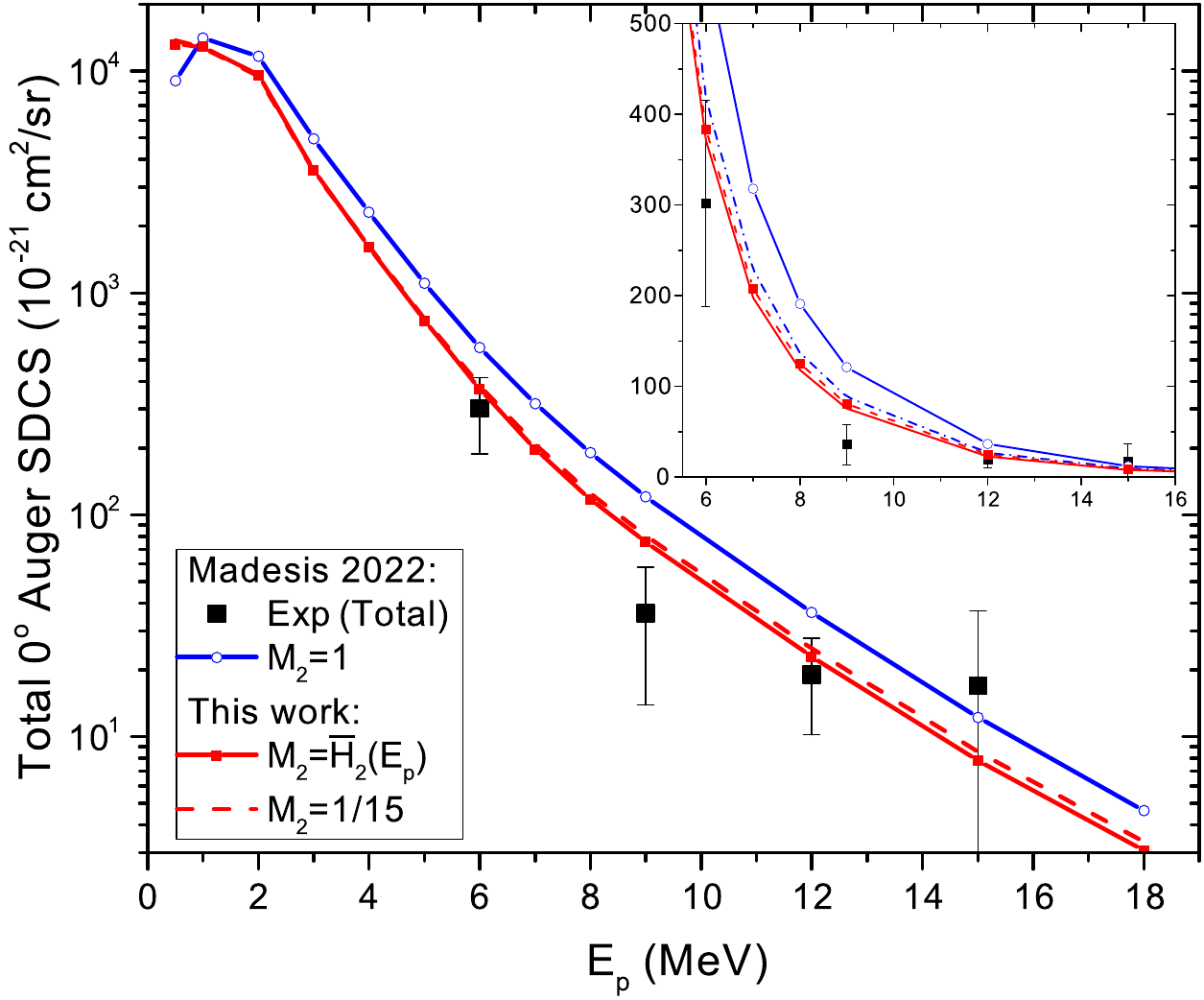}
\caption{Prompt-equivalent $0^\circ$ Auger total (SEC + cascades)
SDCS [Eq.~\eqref{eq:AAD_M2_master}] versus projectile energy $E_p$.
Absolute experimental ZAPS data and the traditional pure $LS$ model
($M_2=1$) are from Ref.~\cite{mad22a}. The three theoretical curves
correspond to the $M_2$ values defined in the text. The inset shows
5.5--16.0~MeV on a linear scale.}
\label{fig:carbon_4P_total}
\end{figure}

Applying these treatments to the total zero-degree SDCS,
Fig.~\ref{fig:carbon_4P_total} shows that the traditional pure $LS$
model overestimates the SDCS across the entire energy range, whereas
the present hybrid model provides improved agreement with experiment.
The corresponding values are compiled in
Table~S2 of the SM.

\begin{figure}[htbp]
\centering
\includegraphics[width=\columnwidth]{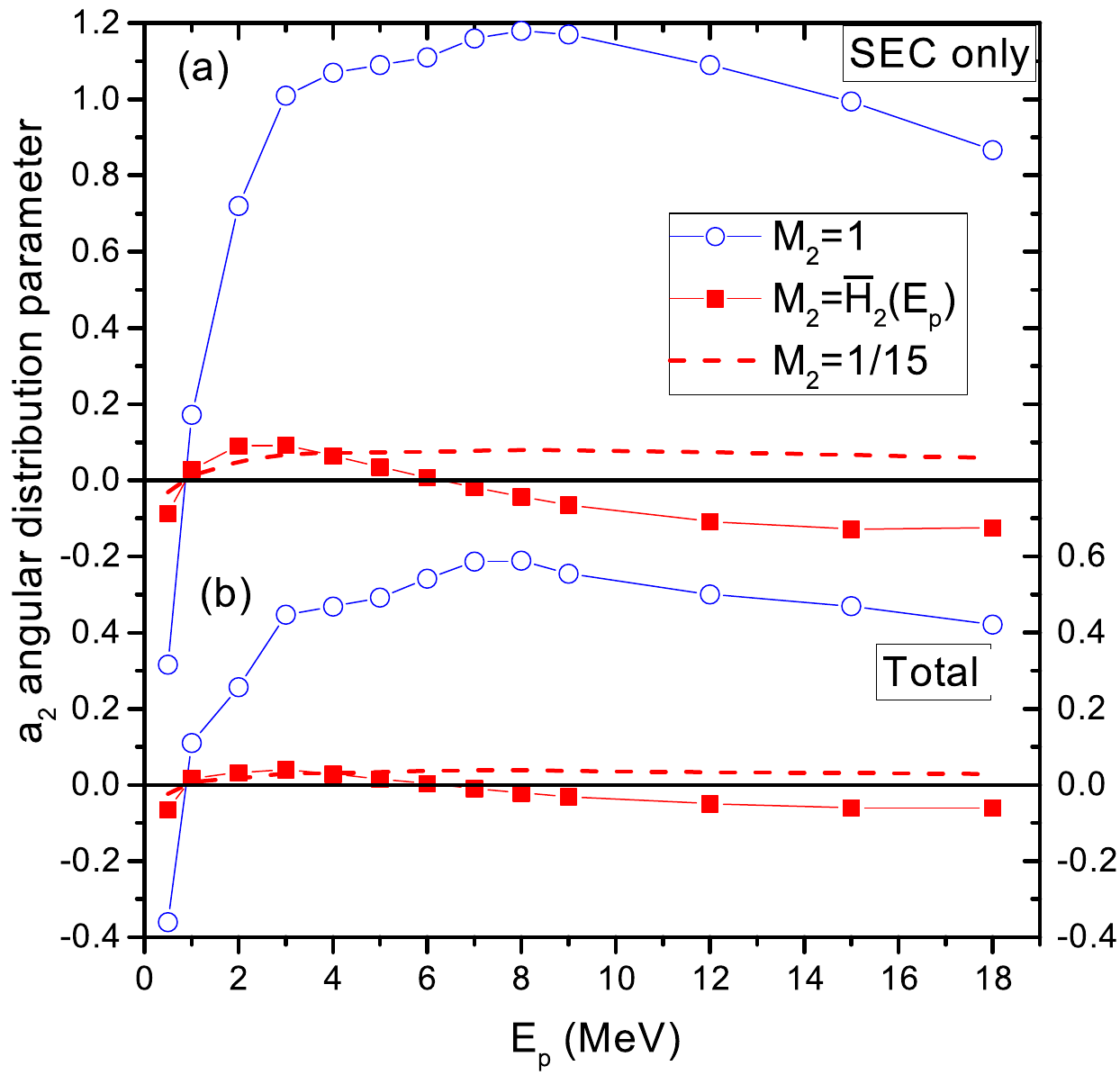}
\caption{Angular distribution parameter $a_2=\alpha_2\mathcal{A}_{20}M_2$ for
(a) SEC only and (b) total (SEC + cascades). The horizontal lines mark
$a_2=0$ (isotropy). The large disparity between $LS$ ($M_2=1$) and present
$LS/LSJ$ [$M_2=\overline{H}_2(E_p)$] results is particularly pronounced for
SEC only (a).}
\label{fig:angular_distribution_a2_total}
\end{figure}

The relatively close agreement among the calculated curves in
Fig.~\ref{fig:carbon_4P_total} results from the isotropic cascade feeding,
which compresses the differences in the absolute SDCS, masking the much
stronger structural divergence between the coupling frameworks. This
difference is more clearly exposed through the anisotropy parameter $a_2$
in Fig.~\ref{fig:angular_distribution_a2_total}.

For SEC only [Fig.~\ref{fig:angular_distribution_a2_total}(a)],
the traditional pure $LS$ model predicts substantially larger
anisotropy than the present hybrid model and its ideal limit.
The distinction persists, although reduced by isotropic cascade
feeding, for the total channel
[Fig.~\ref{fig:angular_distribution_a2_total}(b)].
The corresponding $a_2$ values are compiled in
Tables~S4 and S5 of the SM.

Changing $E_p$ shifts the spatial distribution of the in-flight decays
relative to the spectrometer entrance. This velocity dependence produces
a hybrid-specific zero-crossing:
$\overline{H}_2(E_p)=0$ at $E_p=6.26$~MeV, rendering the unresolved
multiplet isotropic ($a_2=0$). From Eq.~\eqref{eq:H2_overline_def},
this occurs when
$3G_{5/2}\xi_{5/2}-2G_{3/2}\xi_{3/2}=0$.
Crucially, the collision-produced orbital alignment
$\mathcal{A}_{20}[L]$ remains nonzero, as do the individual $J$
component alignments.

The low-energy zero-crossing around $E_p=1$~MeV is common to all three calculations and originates from the collision-produced alignment (i.e., $\sigma^\text{SEC}_{1,|1|}=\sigma^\text{SEC}_{1,0}$; see Table~S1), in contrast to the hybrid zero-crossing at $E_p=6.26$~MeV described above. Above the low-energy crossing and for $E_p<6.26$~MeV, the hybrid model gives positive $a_2$. For $E_p>6.26$~MeV, downstream decay of the highly metastable $J=5/2$ state reduces $G_{5/2}$, shifting the relative fine-structure intensity balance and driving $\overline{H}_2(E_p)$, and hence $a_2$, negative.

At either zero-crossing, the isotropic AAD allows the total cross section to be directly extracted from Eq.~\eqref{eq:AAD_M2_master} without an angular-distribution correction based on the magnetic-substate partial cross sections entering $\mathcal{A}_{20}^{\rm tot}$. This provides a particularly useful point of comparison with collision calculations that accurately predict total production cross sections without resolving their magnetic-substate populations.

The ideal-limit results in Figs.~1--3 isolate the fine-structure projection,
showing that the strong suppression relative to the traditional pure $LS$
result persists without level-dependent solid-angle and Auger-yield effects.

The improved agreement with the measured cross sections in
Fig.~2 demonstrates the importance of retaining the individual fine-structure
decay pathways. Consistent with the small hybrid $a_2$ values in Fig.~3,
the isotropic reference SDCS, $\xi\sigma_{\rm tot}/(4\pi)$, lies close to the
hybrid result over most of the energy range (not shown), becoming identical
at the $E_p=6.26$~MeV zero-crossing.

Although the $J=5/2$ level can support $k=4$ alignment in $LSJ$
coupling, the fine-structure projection gives no $k=4$ contribution
for $L=1$. The improvement with the hybrid model therefore arises
entirely from the pronounced attenuation and sign inversion of $a_2$,
rather than from an expansion of the multipole order.

Unlike the Mehlhorn--Taulbjerg treatment~\cite{mehl80a} for overlapping
fine-structure resonances, the present hybrid framework treats the
well-isolated $J$ channels individually after the fine-structure projection.

Strong changes and even sign inversion of Auger anisotropies have been
predicted for overlapping fine-structure multiplets, where coherent
excitation produces interference between $J$ components~\cite{kab94a}.
In the non-overlapping limit, Kabachnik \textit{et al.} showed that only
the $J=J'$ contributions remain and the fine-structure components add
incoherently~\cite{kab94a}. The present $1s2s2p\,^4\!P_J$ manifold
realizes this isolated-resonance limit. Its suppressed and inverted
unresolved anisotropy therefore arises not from inter-$J$ coherence as
in Ref.~\cite{kab94a}, but directly from projecting the collision-produced
orbital alignment onto the well-isolated fine-structure states.

An even more stringent test of the hybrid framework would be an
$L=2$ state, for which the collision-produced alignment can also contain a
$k=4$ term. One candidate is the carbon $1s2p^2\,^2\!D_J$
($J=3/2,5/2$) state, recently investigated following resonant
transfer-excitation in C$^{4+}(1s^2) + \mathrm{He}$ collisions~\cite{lao22a}.
Its strongly overlapping fine-structure resonances, however, place it outside
the isolated-resonance regime considered here.

Although the isolated-resonance condition does not require metastability,
the metastable $1s2s2p\,^4\!P_J$ manifold provides a particularly suitable
low-$Z_p$ benchmark. A further test is possible for the $1s2s2p\,^4\!P_J$
states produced by $1s\to2p$ excitation with spin exchange in
O$^{5+} + \mathrm{He}$ or F$^{6+} + \mathrm{He}$ collisions. These systems
have been investigated~\cite{zou89b,lee91a}, but, to our
knowledge, no corresponding rigorous collision calculations have
been reported. Because the incident Li-like beam is entirely in its ground
state, this pathway avoids the independent determination of the metastable
beam fraction required for carbon SEC~\cite{mad22a} and thereby removes an
additional source of uncertainty in the absolute cross-section comparison.
Additionally, the different fine-structure lifetimes and Auger yields at
these higher $Z_p$~\cite{ben15a} would test the explicit state-dependent
weighting through $G_J(E_p)$ and $\xi_J$ in the hybrid framework.
However, since spin exchange is involved, a treatment at the nonperturbative AOCC level would require a five-electron calculation, which is presently beyond practical computational capabilities~\cite{dub26a}.
Perturbative approaches have also been applied to projectile excitation~\cite{sur08a,lao24a}, but to our knowledge without spin exchange.

In conclusion, we have presented a general hybrid $LS/LSJ$ framework for
atomic decay angular distributions of fine-structure multiplets with isolated
components.
Within the two-step model, state production is treated in $LS$
coupling, and the resulting orbital alignment is projected onto the individual
fine-structure states by angular-momentum recoupling of the $LS$ magnetic-sublevel
populations before their decay angular distributions are computed in $LSJ$
coupling.
Applied to the multi-open-shell $1s2s2p\,^4\!P_J$ manifold populated via SEC
in $\text{C}^{4+}(1s2s\,^3\!S) + \text{He}$ collisions, the framework
reveals pronounced suppression and inversion of the anisotropy relative
to the traditional pure $LS$ treatment and improves agreement with absolute
experimental data using the \textit{same} $LS$ production cross
sections~\cite{mad22a}.

For the isolated, non-overlapping $^4\!P_J$ levels, the contributions
add incoherently, demonstrating that inversion of the unresolved anisotropy
does not require inter-$J$ coherence.
This incoherent hybrid $LS/LSJ$ treatment applies more generally when
$\Delta E_\text{FS}\gg\Gamma$, including multi-open-shell systems.
The significance of this distinction extends beyond the present system.
Fine structure can qualitatively alter the decay angular distribution even
when the collision dynamics remain well described in $LS$ coupling.

\begin{acknowledgments}
\paragraph{Acknowledgments.} The author is grateful to Peter Rakitzis, Alain Dubois and Manolis Benis for valuable discussions, and thanks Stefanos Nanos for constructive comments on the manuscript. Special thanks are due to Andrey Surzhykov for stimulating correspondence that initiated this work.
\end{acknowledgments}



\clearpage
\appendix
\setcounter{secnumdepth}{1}
\section{Isolated-resonance condition}
\label{apx:sec_IRA}

For the $1s2s2p\,^4\!P_J$ manifold considered here, the condition
$\Delta E_{\rm FS}\gg\Gamma$ corresponds to the isolated-resonance
approximation (IRA). Following Ref.~\cite{kab94a}, for a pair of
fine-structure levels we define
$\Gamma_{JJ'}=(\Gamma_J+\Gamma_{J'})/2$ and
$\epsilon_{JJ'}=|\Delta E_{JJ'}|/\Gamma_{JJ'}$.
As shown in Table~\ref{tb:hybridcriterion}, all three fine-structure
pairs have $\epsilon_{JJ'}\gg1$, strongly confirming the validity of
the IRA for the $^4\!P_J$ manifold.

\begin{table}[h]
\caption{\label{tb:hybridcriterion}
Fine-structure energy separations, pairwise natural widths
$\Gamma_{JJ'}=(\Gamma_J+\Gamma_{J'})/2$, and corresponding
isolated-resonance ratios for C$^{3+}(1s2s2p\,^4\!P_J)$.
The energies are from Refs.~\cite{yer17a,*yer17b}, and the natural
widths are from Ref.~\cite{ben15a}.}
\begin{ruledtabular}
\begin{tabular}{lccc}
$J,J^\prime$ & $1/2,3/2$ & $1/2,5/2$ & $3/2,5/2$ \\
\hline
$|\Delta E_{JJ^\prime}|$ (eV)
& 5.44228E-4 & 1.25172E-2 & 1.19730E-2 \\
$\Gamma_{JJ^\prime}$ (eV)
& 158.35E-9 & 114.71E-9 & 49.06E-9 \\
$\epsilon_{JJ^\prime}
=|\Delta E_{JJ^\prime}|/\Gamma_{JJ^\prime}$
& 3.437E03 & 1.091E05 & 2.440E05 \\
\end{tabular}
\end{ruledtabular}
\end{table}

\section{Generalized Auger Angular Distributions}

In the emitter frame, with the quantization $z$ axis along the
collision axis, the generalized AAD is axially symmetric. The corresponding
Auger SDCS is traditionally expanded in Legendre polynomials,
$P_k(\cos\theta)$, as:
\begin{equation}
\frac{d\sigma_A}{d\Omega_0}(\theta) = \xi\,\frac{\sigma}{4\pi} \left[ 1 + \sum_{k} a_k P_k(\cos \theta) \right],
\label{eq:aad_expansion}
\end{equation}
where $\sigma_A$ represents the solid-angle-integrated Auger cross section,
and $\theta$ is the electron emission angle relative to the beam axis.
The Auger cross section is related to the primary production cross
section $\sigma$ via the state-specific Auger yield $\xi$ (i.e., the
non-radiative branching ratio), such that $\sigma_A = \xi \sigma$.
$d\Omega_0$ denotes the effective point-source solid angle for prompt Auger
emission.

For the axially symmetric collision geometry considered here, the magnetic
orbital substates $M_L$ and $-M_L$ are populated equally, so that the
intermediate state may be aligned but not oriented. Consequently, the
summation index $k$ in Eq.~\eqref{eq:aad_expansion} is restricted to even
integers ($k=2,4,\ldots$). The maximum rank is further constrained by the
angular momentum of the aligned state, with $k\leq2L$ in $LS$ coupling and
$k\leq2J$ in $LSJ$ coupling~\cite{kab94a}.

Within the two-step model, the angular distribution coefficients $a_k$
factorize into the alignment and intrinsic anisotropy parameters:
\begin{align}
    a_{k}&[L_0S_0J_0\rightarrow LSJ\rightarrow L_fS_fJ_f] \nonumber\\
    &= \alpha_{k}[LSJ\rightarrow L_fS_fJ_f] \cdot
    \mathcal{A}_{k0}[L_0S_0J_0\rightarrow LSJ].
    \label{eq:twostep_factorization}
\end{align}
Here, $L$, $S$, and $J$ denote the orbital, spin, and total angular momentum, respectively, with subscripts $0$ and $f$ denoting the
initial and final states; quantities without subscripts refer to the
intermediate state. At each stage, $\mathbf{J}=\mathbf{L}+\mathbf{S}$.
The parameters $\mathcal{A}_{k0}$ and $\alpha_k$ are the alignment and
intrinsic anisotropy parameters, respectively~\cite{kab94a}.

The alignment parameter $\mathcal{A}_{k0}$ characterizes the production of
the excited state in the first step and is determined by the corresponding
magnetic-substate-resolved production cross sections, which generally depend
on the collision energy $E_p$. Conversely, the intrinsic anisotropy parameter
$\alpha_k$ characterizes the subsequent Auger emission in the second step,
depends only on the atomic structure and decay channels, and is independent of
$E_p$.

\section{The Alignment Parameter $\mathcal{A}_{k0}[L]$}

Following Kabachnik \textit{et al.}~\cite{kab94a}, for $LS$-coupled
production the orbital alignment parameter $\mathcal{A}_{k0}[L]$ is defined in
terms of the statistical tensors as
$\mathcal{A}_{k0}[L]=\rho_{k0}(L,L)/\rho_{00}(L,L)$.
Using the general expression for $\rho_{k0}$ given by Berezhko and
Kabachnik~\cite{ber77a}, and expressing the magnetic-substate populations
in terms of the partial production cross sections
$\sigma(L,|M_L|)$, this becomes:
{\small
\begin{equation}
\mathcal{A}_{k0}[L] =
\frac{\hat{L}\,\hat{k}}{\sigma[L]}
\sum_{M_L=-L}^{L}(-1)^{L-M_L}
\begin{pmatrix}
L & L & k \\
M_L & -M_L & 0
\end{pmatrix}
\sigma(L,|M_L|),
\label{eq:calAk0L_def}
\end{equation}
}%
where $\hat{X} \equiv \sqrt{2X+1}$ for any angular momentum or tensor rank
variable. The corresponding total production cross section $\sigma[L]$
normalizing the tensor is given by:
\begin{equation}
\sigma[L] = \sum_{M_L=-L}^{L} \sigma(L,|M_L|).
\label{eq:sigmaL}
\end{equation}

For low-$Z_p$ emitters investigated by
ZAPS ($Z_p=2\text{--}10$), the fast collision dynamics are dominated by the
electrostatic Coulomb interaction and are well described by $LS$-coupled
production. Consequently, the alignment parameter in ZAPS is calculated
from the magnetic-substate production cross sections using
Eq.~\eqref{eq:calAk0L_def}. An $LSJ$ treatment of the collision dynamics may
be necessary only when spin-dependent relativistic interactions become
significant in production, as can occur for high-$E_p$, high-$Z_p$
emitters~\cite{sur08a}.

\section{Intrinsic Anisotropy Parameters and the Single Partial-Wave Limit}

While the alignment parameters quantify the state production during the
collision, the intrinsic anisotropy parameters $\alpha_k$ govern the angular
distributions of the subsequent decay. The Auger electron may, in general,
be emitted in different continuum partial waves characterized by its orbital
and total angular momenta $(l,j)$. When several partial waves contribute,
$\alpha_k$ depends on their complex-valued radial matrix elements and
scattering phase shifts~\cite{kab84a}, requiring detailed numerical modeling of the
continuum wavefunctions~\cite{che92a}.

For a wide class of autoionizing states, however, angular momentum
and parity selection rules restrict the decay to a single $(l,j)$ partial
wave. In this single-partial-wave limit, the radial matrix element and
scattering phase cancel from $\alpha_k$, leaving $\alpha_k$ determined entirely by angular-momentum coupling~\cite{che92a}.

For the $1s2s2p\,^4\!P_J$ manifold considered here, the Auger decay
terminates in the closed-shell $1s^2\,^1\!S_0$ final ionic state
($J_f=0$). Total angular momentum conservation then requires $j=J$.
For a given $j$, the electron orbital angular momenta $l=j\pm1/2$
have opposite parity, so that parity conservation selects a unique $l$.
Each $J$ level therefore decays through a single $(l,j)$ partial wave.
Rather than retaining the traditional continuum representation that
explicitly depends on both the ejected electron's orbital $l$ and total
angular momentum $j$~\cite{che92a,bal00b}
(see Eq.~(S69) in the SM), we apply the Wigner
$3j$-$6j$ reduction identity~\cite{des63a} to eliminate the dependence
on $l$ (see Eq.~(S68) in the SM). This reduces
the angular coupling matrix to a total-$J$ expression:
\begin{equation}
\alpha_k^{(j=J)}[J,J_f=0] =(-1)^{J-\frac{1}{2}+k}\,
\hat{J}\,\hat{k}
\begin{pmatrix}
J & J & k \\
\frac{1}{2} & -\frac{1}{2} & 0
\end{pmatrix}.
\label{eq:alpha_J_unreduced}
\end{equation}
In this form, the radial matrix element and scattering phase have canceled,
while the explicit dependence on the electron orbital angular momentum $l$
has been eliminated by the angular-momentum reduction.

Correspondingly, in $LS$ coupling, the intrinsic anisotropy parameter
$\alpha^{(l)}_k[L,L_f]$ is evaluated in terms of  orbital angular
momenta~\cite{kab94a}. For single $l$-wave emission with $L_f=0$, orbital
angular momentum conservation requires $l=L$, reducing the general
expression to:
\begin{equation}
\alpha_k^{(l=L)}[L,L_f=0]
=
(-1)^L\,
\hat{L}\,\hat{k}
\begin{pmatrix}
L & L & k \\
0 & 0 & 0
\end{pmatrix}.
\label{eq:alphak_singlel}
\end{equation}

While $LS$-coupled production describes the collision-produced
orbital alignment in low-$Z_p$ systems, $LS$ coupling may be insufficient
to describe the subsequent autoionization. For the
$1s2s2p\,^4\!P_J$ manifold considered here, spin conservation forbids Coulomb
autoionization to the $1s^2\,^1\!S_0$ ground state in pure $LS$ coupling.
Autoionization is therefore enabled by relativistic Breit-Pauli
interactions~\cite{dav89a}, requiring the explicit treatment of the individual $J$ fine-structure
channels. This motivates preserving
the collision-produced orbital alignment while imposing the state-specific
$LSJ$ decay constraints.

\section{Cascade Contribution to the Total Anisotropy}
\label{app:cascade_a2}

For the present $1s2s2p\,^4\!P_J$ manifold, the alignment entering the
total (SEC + cascades) angular distribution follows from
Eq.~\eqref{eq:calAk0L_def}. With $\alpha_2=-\sqrt{2}$,
\begin{align}
\alpha_2\,\mathcal{A}_{20}^\text{tot} &= (-\sqrt{2})\cdot\left[(\sqrt{2})\,\left(\frac{\sigma^\text{tot}_{1,|1|} - \sigma^\text{tot}_{1,0}}{\sigma^\text{tot}}\right) \right].\nonumber
\intertext{Separating the SEC and cascade contributions gives:}
\alpha_2\,\mathcal{A}_{20}^\text{SEC+casc}&= 2\,\left[\frac{\sigma^\text{SEC}_{1,0}+\sigma^\text{casc}_{1,0} - (\sigma^\text{SEC}_{1,|1|}+\sigma^\text{casc}_{1,|1|})}{\sigma^\text{tot}}\right], \nonumber \\
&= 2\,\left[\left(\frac{\sigma^\text{SEC}_{1,0} - \sigma^\text{SEC}_{1,|1|}}{\sigma^\text{SEC}}\right) \left(\frac{\sigma^\text{SEC}}{\sigma^\text{tot}}\right)\right],\nonumber\\
& = \alpha_2\,\mathcal{A}_{20}^\text{SEC}\,\left(\frac{\sigma^\text{SEC}}{\sigma^\text{tot}}\right),\label{eq:a2calA20SECtimesratio}
\end{align}
where $\sigma^\text{tot}_{L,M_L}=\sigma^\text{SEC}_{L,M_L}+\sigma^\text{casc}_{L,M_L}$.
Because the cascade contribution is isotropic~\cite{zou20a}
($\sigma^{\rm casc}_{1,0}=\sigma^{\rm casc}_{1,|1|}$), it cancels from
the alignment numerator but remains in the denominator
($\sigma^{\rm tot}=\sigma^{\rm SEC}+3\sigma^{\rm casc}_{1,0}$),
thereby diluting the SEC alignment by the factor
$\sigma^{\rm SEC}/\sigma^{\rm tot}$. In Madesis \textit{et al.}~\cite{mad22a},
the partial cross sections for SEC and cascades were computed in pure $LS$
coupling by AOCC calculations; their values are listed in
Table~S1 of the SM.


\clearpage
\onecolumngrid
\renewcommand{\appendixname}{}
\setcounter{section}{0}
\setcounter{subsection}{0}
\setcounter{subsubsection}{0}
\setcounter{equation}{0}
\setcounter{figure}{0}
\setcounter{table}{0}
\setcounter{page}{1}
\setcounter{secnumdepth}{3}
\renewcommand{\thesection}{\Alph{section}}
\renewcommand{\thesubsection}{\thesection.\arabic{subsection}}
\renewcommand{\thesubsubsection}{\thesubsection.\arabic{subsubsection}}

\makeatletter
\renewcommand{\p@subsection}{}
\renewcommand{\p@subsubsection}{}
\makeatother

\renewcommand{\theequation}{S\arabic{equation}}
\renewcommand{\thefigure}{S\arabic{figure}}
\renewcommand{\thetable}{S\arabic{table}}
\renewcommand{\thepage}{S\arabic{page}}
\renewcommand{\theHequation}{S\arabic{equation}}
\renewcommand{\theHfigure}{S\arabic{figure}}
\renewcommand{\theHtable}{S\arabic{table}}

\vspace*{3mm}
\begin{center}
{\bf Supplemental Material for}\\[1ex]
{\bf ``Strong Angular-Momentum-Coupling Dependence \\of Atomic Decay
Angular Distributions in a Hybrid $\mathbf{LS/LSJ}$ Framework''}\\[1ex]
T. J. M. Zouros\\
Department of Physics, University of Crete, GR-70013 Heraklion, Greece\\[1ex]
\mbox{}\par
\end{center}

This supplemental material provides the detailed mathematical formulations,
atomic parameters, and numerical treatments underlying the $LS/LSJ$ hybrid
coupling model for the carbon
$1s2s2p\,^4\!P_J \rightarrow 1s^2\ ^1\!S_0$ Auger transition. We consider
$\text{C}^{4+}(1s2s\,^3\!S)+\text{He}
\rightarrow \text{C}^{3+}(1s2s2p\ ^4\!P_J)$ collisions, investigated
theoretically and experimentally by zero-degree Auger projectile spectroscopy
(ZAPS) in Ref.~\cite{mad22a}. The $^4\!P_J$ levels are populated directly by
$2p$ single electron capture (SEC) and indirectly through cascades from
higher-lying states~\cite{zou20a}.

The SEC magnetic-substate partial cross sections $\sigma(L,M_L)$  were
calculated as a function of projectile energy $E_p$ using three-electron
atomic orbital close-coupling (3eAOCC) calculations, while the cascade
contribution was evaluated using one-electron AOCC (1eAOCC) calculations. These partial cross sections were used to compute the alignment parameter
$\mathcal{A}_{20}[L=1]$ using Eq.~(C1).
Absolute Auger single differential cross sections (SDCS) were measured at
$\theta=0^\circ$. The cross sections used throughout this work are those of
Ref.~\cite{mad22a}.

\section{Detailed evaluation of the pure $LS$ and hybrid $LS/LSJ$ treatments
for the zero-degree Auger angular distributions}\label{apx:sec_aad_derivation}

\subsection{Foundational Cross-Section Definitions and Notation}
\label{apx:sec_cs_defs}

We establish the total, SEC, and cascade cross sections for the
$1s2s2p\,^4\!P_J$ ($L=1, S=3/2$) manifold in terms of their
magnetic-substate partial cross sections.
For unpolarized initial states and axial symmetry along the beam axis,
the positive and negative magnetic substates are populated equally,
$\sigma(L,M_L)=\sigma(L,-M_L)\equiv\sigma(L,|M_L|)$.
The total cross sections are then defined by summing these partial cross
sections [see Eq.~(C2)]:
\begin{align}
\sigma^{\text{tot}} &= \sigma^{\text{SEC}} + \sigma^{\text{casc}}, \label{apx:eq_sigmatot} \\
\sigma^{\text{SEC}} &= \sigma^{\text{SEC}}(1,0) + 2\sigma^{\text{SEC}}(1,|1|), \label{apx:eq_sigmaSEC} \\
\sigma^{\text{casc}} &= \sigma^{\text{casc}}(1,0) + 2\sigma^{\text{casc}}(1,|1|). \label{apx:eq_sigmacasc}
\end{align}

Because indirect cascade contributions are isotropic, their partial cross
sections are equal,
$\sigma^{\text{casc}}(1,|1|) = \sigma^{\text{casc}}(1,0)$.
The total substate components therefore become:
\begin{align}
\sigma^{\text{tot}}(1,0) &= \sigma^{\text{SEC}}(1,0) + \sigma^{\text{casc}}(1,0), \label{apx:eq_sigma10tot} \\
\sigma^{\text{tot}}(1,|1|) &= \sigma^{\text{SEC}}(1,|1|) + \sigma^{\text{casc}}(1,0), \label{apx:eq_sigma11tot} \\
\sigma^{\text{tot}} &= \sigma^{\text{SEC}}(1,0) + 2\sigma^{\text{SEC}}(1,|1|) + 3\sigma^{\text{casc}}(1,0). \label{apx:eq_sigma1tot}
\end{align}

\subsection{Evaluation of the Zero-Degree Pure $LS$ Coupling Baseline}
\label{apx:sub_eval_LS}

For the pure $LS$ treatment, the prompt-equivalent zero-degree
($\theta=0^\circ$) Auger angular distribution (AAD) for the unresolved
$1s2s2p\,^4\!P$ ($L=1$) multiplet is [see Eq.~(14) with $M_2=1$]
\begin{align}
\frac{d\sigma_A^{LS}[\,^4\!P]}
     {\overline{G}_\tau\,d\Omega_0}(0^\circ)
&=
\overline{\xi}\,\frac{\sigma[\,^4\!P]}{4\pi}
\left(1+\alpha_2\mathcal{A}_{20}\right).
\end{align}
Here, $\mathcal{A}_{20}$ is the $L=1$ orbital alignment parameter defined in
Eq.~(C1), and $\alpha_2$ is the corresponding pure-$LS$
intrinsic anisotropy parameter given in Eq.~(D2).

For $L=1$, expanding Eq.~(C1) over the individual
$M_L=\{-1,0,1\}$ magnetic substates gives
\begin{align}
\mathcal{A}_{20}[L=1]
&=
\frac{\hat{1}\cdot\hat{2}}{\sigma}
\sum_{M_L=-1}^{1}
(-1)^{1-M_L}
\begin{pmatrix}
1 & 1 & 2\\
M_L & -M_L & 0
\end{pmatrix}
\sigma(1,|M_L|)
\nonumber\\
&=
\frac{\sqrt{15}}{\sigma}
\left[
(-1)^2
\begin{pmatrix}
1&1&2\\
-1&1&0
\end{pmatrix}
\sigma(1,|-1|)
+
(-1)^1
\begin{pmatrix}
1&1&2\\
0&0&0
\end{pmatrix}
\sigma(1,0)
\right.
\nonumber\\
&\hspace{3.0cm}\left.
+
(-1)^0
\begin{pmatrix}
1&1&2\\
1&-1&0
\end{pmatrix}
\sigma(1,|1|)
\right].
\end{align}
Substituting the analytical values of the Wigner $3j$ symbols gives
\begin{align}
\mathcal{A}_{20}[L=1]
=
\frac{1}{\sigma}
\left[
\frac{\sqrt{2}}{2}\sigma(1,|-1|)
-\sqrt{2}\sigma(1,0)
+\frac{\sqrt{2}}{2}\sigma(1,|1|)
\right].
\end{align}
Using axial symmetry,
$\sigma(1,|-1|)=\sigma(1,|1|)$, this reduces to
\begin{align}
\mathcal{A}_{20}[L=1]
=
\sqrt{2}
\left[
\frac{\sigma(1,|1|)-\sigma(1,0)}{\sigma}
\right].\label{apx:eq_calA20Leq1_final_reduced}
\end{align}

For single partial-wave emission with $l=L=1$ and $L_f=0$, the pure-$LS$
intrinsic anisotropy parameter is
\begin{align}
\alpha_2
&=
\alpha_2^{(l=1)}[L=1,L_f=0]
=
(-1)^1\hat{1}\cdot\hat{2}
\begin{pmatrix}
1&1&2\\
0&0&0
\end{pmatrix}
\nonumber\\
&=
(-1)\sqrt{3}\sqrt{5}
\left(\frac{\sqrt{30}}{15}\right)
=
-\sqrt{2}.\label{apx:eq_alpha2_Leq1_val}
\end{align}
We therefore obtain the well-known~\cite[see Table 1]{mehl80a} pure-$LS$
angular distribution parameter $a_2$ for $L=1$,
\begin{align}
a_2 = \alpha_2\mathcal{A}_{20}
=
-2\left[
\frac{\sigma(1,|1|)-\sigma(1,0)}{\sigma}
\right]
=
2\left[
\frac{\sigma(1,0)-\sigma(1,|1|)}{\sigma}
\right].
\end{align}
Substitution into the zero-degree pure-$LS$ distribution, with
$\sigma=\sigma(1,0)+2\sigma(1,|1|)$, gives
\begin{align}
\frac{d\sigma_A^{LS}[\,^4\!P]}
     {\overline{G}_\tau\,d\Omega_0}(0^\circ)
&=
\overline{\xi}\,\frac{\sigma}{4\pi}
\left[
1-2\frac{\sigma(1,|1|)-\sigma(1,0)}{\sigma}
\right]
\nonumber\\
&=
3\,\overline{\xi}\,
\frac{\sigma(1,0)}{4\pi}.
\end{align}
Thus, in pure $LS$ coupling, the zero-degree AAD exhibits the familiar
$M_L=0$ filtering: only the $M_L=0$ partial cross section contributes
at $\theta=0^\circ$.

The Mehlhorn--Taulbjerg formalism leads to an AAD of analogous form with
an additional de-alignment factor. The corresponding treatment, applicable
to the overlapping fine-structure regime, is discussed separately in
Sec.~\ref{apx:subsub_Dk_parameter}.

\subsection{Evaluation of the Zero-Degree $LS/LSJ$ Hybrid Coupling Treatment}
\label{apx:sub_eval_LSJ}

For the hybrid $LS/LSJ$ treatment, the unresolved $^4\!P$ multiplet SDCS
is obtained by summing the active fine-structure channels
$J=1/2,3/2,$ and $5/2$ in
Eq.~(5), as performed in Eqs.~(6)-(7). At zero degrees
($\theta=0^\circ$), relative to the prompt point-source solid angle
element $d\Omega_0$, we introduce the level-intensity shorthand
$X_A[^4\!P_J]\equiv G[^4\!P_J]\,\xi[^4\!P_J]\,\sigma[^4\!P_J]$,
or simply $X_A[J]\equiv G_J\,\xi_J\,\sigma[J]$:
\begin{equation}
\frac{d\sigma^{\text{hybrid}}_A[^4\!P]}{d\Omega_0}(0^\circ) = \frac{X_A[1/2]}{4\pi} \left( 1 \right) + \frac{X_A[3/2]}{4\pi} \left( 1 + \alpha_2^{(3/2)} \mathcal{A}^{\text{proj}}_{20}[3/2] \right) + \frac{X_A[5/2]}{4\pi} \left( 1 + \alpha_2^{(5/2)} \mathcal{A}^{\text{proj}}_{20}[5/2] + \alpha_4^{(5/2)} \mathcal{A}^{\text{proj}}_{40}[5/2] \right),
\label{eq:hybrid_sum_general_with_k4}
\end{equation}
where the Legendre polynomials reduce identically to unity,
$P_k(1)=1$. We drop the explicit $[^4\!P]$ state labels hereafter for
brevity. To evaluate the individual channel contributions, we apply the
geometrical projection of Eq.~(1) to the
quadrupole ($k=2$) and hexadecapole ($k=4$) ranks of the
$1s2s2p\,^4\!P_J$ manifold ($L=1,S=3/2$):
\begin{equation}
\mathcal{A}_{k0}^{\rm proj}[L=1,S=3/2,J] = (-1)^{1+3/2+J+k}\hat{1}\cdot\hat{J} \begin{Bmatrix} 1 & J & 3/2 \\ J & 1 & k \end{Bmatrix}\mathcal{A}_{k0},\label{apx:eq_calAk0_proj_def}
\end{equation}
where $\mathcal{A}_{k0}$ is the pure-$LS$ orbital alignment parameter
defined in Eq.~(C1).
Therefore, for $L=1$ and $S=3/2$ we have:
\begin{align}
\mathcal{A}^{\text{proj}}_{20}[1/2] &= (-1)^{1 + \frac{3}{2} + \frac{1}{2} + 2} \, \hat{1} \cdot \hat{\frac{1}{2}} \begin{Bmatrix} 1 & \frac{1}{2} & \frac{3}{2} \\ \frac{1}{2} & 1 & 2 \end{Bmatrix} \mathcal{A}_{20} = 0, \\
\mathcal{A}^{\text{proj}}_{20}[3/2] &= (-1)^{1 + \frac{3}{2} + \frac{3}{2} + 2} \, \hat{1}\cdot \hat{\frac{3}{2}} \begin{Bmatrix} 1 & \frac{3}{2} & \frac{3}{2} \\ \frac{3}{2} & 1 & 2 \end{Bmatrix} \mathcal{A}_{20} = -\frac{2\sqrt{2}}{5} \mathcal{A}_{20}, \\
\mathcal{A}^{\text{proj}}_{20}[5/2] &= (-1)^{1 + \frac{3}{2} + \frac{5}{2} + 2} \, \hat{1} \cdot \hat{\frac{5}{2}} \begin{Bmatrix} 1 & \frac{5}{2} & \frac{3}{2} \\ \frac{5}{2} & 1 & 2 \end{Bmatrix} \mathcal{A}_{20} = \frac{\sqrt{7}}{5} \mathcal{A}_{20}.
\end{align}
The $6j$ symbol for $J=1/2$ vanishes because the triangular condition
$\Delta(j_4,j_2,j_6)\rightarrow(1/2,1/2,2)$ is not satisfied,
consistent with the fact that a $J=1/2$ state cannot support a
quadrupole alignment tensor ($k=2$).

For the hexadecapole components, the triangle condition
$\Delta(J,J,k)$ restricts nonzero contributions to $k\leq2J$, so that
$\mathcal{A}^{\text{proj}}_{40}[1/2]=
\mathcal{A}^{\text{proj}}_{40}[3/2]=0$. For $J=5/2$,
\begin{equation}
\mathcal{A}^{\text{proj}}_{40}[5/2] = (-1)^{1 + \frac{3}{2} + \frac{5}{2} + 4} \, \hat{1} \cdot\hat{\frac{5}{2}} \begin{Bmatrix} 1 & \frac{5}{2} & \frac{3}{2} \\ \frac{5}{2} & 1 & 4 \end{Bmatrix} \mathcal{A}_{40} = 0.
\label{eq:k4_projection_equation}
\end{equation}
Here the $6j$ symbol also vanishes because the internal triangular
condition $\Delta(L,L,k)\rightarrow(1,1,4)$ is not satisfied. Thus,
although $J=5/2$ can in principle support a rank-4 tensor, projection
from the collision-produced $L=1$ alignment eliminates the $k=4$
contribution.

To evaluate the level-specific intrinsic anisotropy parameters
$\alpha_k^{(J)}$, we use the reduced single partial-wave expression from
Eq.~(D1) for the closed-shell final ionic core
($J_f=0$ and $j=J$):
\begin{equation}
\alpha_k^{(J)}[J,J_f=0]
=
(-1)^{J-\frac{1}{2}+k}\,
\hat{J}\cdot\hat{k}
\begin{pmatrix}
J & J & k \\
\frac{1}{2} & -\frac{1}{2} & 0
\end{pmatrix}.
\label{eq:alpha_J_eval}
\end{equation}

Evaluating Eq.~\eqref{eq:alpha_J_eval} for the active channels gives the
quadrupole ($k=2$) and hexadecapole ($k=4$) coefficients:
\begin{align}
\alpha_2^{(1/2)}
&=0,
\label{eq:alpha2_J12}
\\
\alpha_2^{(3/2)}
&=
(-1)^{\frac{3}{2}-\frac{1}{2}+2}
\hat{\frac{3}{2}}\cdot\hat{2}
\begin{pmatrix}
\frac{3}{2} & \frac{3}{2} & 2 \\
\frac{1}{2} & -\frac{1}{2} & 0
\end{pmatrix}
=-1,
\label{eq:alpha2_J32}
\\
\alpha_2^{(5/2)}
&=
(-1)^{\frac{5}{2}-\frac{1}{2}+2}
\hat{\frac{5}{2}}\cdot\hat{2}
\begin{pmatrix}
\frac{5}{2} & \frac{5}{2} & 2 \\
\frac{1}{2} & -\frac{1}{2} & 0
\end{pmatrix}
=
-\frac{2\sqrt{14}}{7},
\label{eq:alpha2_J52}
\\
\alpha_4^{(3/2)}
&=0,
\label{eq:alpha4_J32}
\\
\alpha_4^{(5/2)}
&=
(-1)^{\frac{5}{2}-\frac{1}{2}+4}
\hat{\frac{5}{2}}\cdot\hat{4}
\begin{pmatrix}
\frac{5}{2} & \frac{5}{2} & 4 \\
\frac{1}{2} & -\frac{1}{2} & 0
\end{pmatrix}
=
\frac{\sqrt{42}}{7}.
\label{eq:alpha4_J52}
\end{align}
Taking the products of these anisotropy parameters with their
corresponding projected alignment tensors gives
\begin{align}
\alpha_2^{(1/2)} \mathcal{A}^{\text{proj}}_{20}[1/2] &= 0, \\
\alpha_2^{(3/2)} \mathcal{A}^{\text{proj}}_{20}[3/2] &= (-1) \left(-\frac{2\sqrt{2}}{5} \mathcal{A}_{20}\right) = +\frac{2\sqrt{2}}{5} \mathcal{A}_{20}, \label{eq:alpha232A20pro32}\\
\alpha_2^{(5/2)} \mathcal{A}^{\text{proj}}_{20}[5/2] &= \left(-\frac{2\sqrt{14}}{7}\right) \left(\frac{\sqrt{7}}{5} \mathcal{A}_{20}\right) = -\frac{2\sqrt{2}}{5} \mathcal{A}_{20}, \label{eq:alpha252A20pro52}\\
\alpha_4^{(5/2)} \mathcal{A}^{\text{proj}}_{40}[5/2] &= \left(\frac{\sqrt{42}}{7}\right) \left( 0 \right) = 0.
\end{align}
Substituting these channel products into
Eq.~\eqref{eq:hybrid_sum_general_with_k4} gives
\begin{equation}
\frac{d\sigma^{\text{hybrid}}_A[^4\!P]}{d\Omega_0}(0^\circ) = \frac{X_A[1/2]}{4\pi} + \frac{X_A[3/2]}{4\pi}\left(1 + \frac{2\sqrt{2}}{5} \mathcal{A}_{20}\right) + \frac{X_A[5/2]}{4\pi}\left(1 - \frac{2\sqrt{2}}{5} \mathcal{A}_{20}\right).
\label{eq:hybrid_expanded_lines}
\end{equation}
Regrouping the scalar and anisotropic contributions gives
\begin{equation}
\frac{d\sigma^{\text{hybrid}}_A[^4\!P]}{d\Omega_0}(0^\circ) = \frac{\sum_J X_A[J]}{4\pi} \Bigl( 1 + \alpha_2\, \mathcal{A}_{20}\,\overline{H}_2 \Bigr),
\label{eq:hybrid_regrouped}
\end{equation}
where $\alpha_2$ and $\mathcal{A}_{20}$ are given in
Eqs.~\eqref{apx:eq_alpha2_Leq1_val} and
\eqref{apx:eq_calA20Leq1_final_reduced}, respectively, and the effective hybrid
projection factor $\overline{H}_2$ is
\begin{equation}
\overline{H}_2 \equiv \frac{2}{5} \left( \frac{X_A[5/2] - X_A[3/2]}{\sum_J X_A[J]} \right).
\label{eq:H2_isolated_general}
\end{equation}

To evaluate the channel-intensity fractions explicitly, we use the
statistical relation between the fine-structure cross sections
$\sigma[J]$ and the total uncoupled multiplet cross section $\sigma[L]$
(see Sec.~\ref{apx:sec_LStoLSJ_cs_transformations}):
\begin{equation}
\sigma[J] = \frac{\hat{J}^2}{\sum_J \hat{J}^2}\sigma[L] = \frac{\hat{J}^2}{\hat{L}^2\,\hat{S}^2 }\sigma[L],
\label{apx:eq_sigmaJ_to_sigmaL}
\end{equation}
giving for $X_A[J]$ and its sum
\begin{align}
X_A[J] &= G_J \xi_J \sigma[J]  = G_J \xi_J\,\frac{\hat{J}^2}{\hat{L}^2\,\hat{S}^2}\sigma[L], \label{apx:eq_X_A_def} \\
\sum_J X_A[J] &= \sigma[L]\,\left(\frac{\sum_J \hat{J}^2 G_J \xi_J}{\sum_J \hat{J}^2}\right) =\sigma[L]\,\left(\frac{\sum_J \hat{J}^2 G_J \xi_J}{\sum_J \hat{J}^2\xi_J}\right)\left(\frac{\sum_J \hat{J}^2\xi_J}{\sum_J \hat{J}^2}\right) = \overline{G}_\tau\cdot\overline{\xi}\cdot\sigma[L], \label{apx:eq_sum_X_expand}
\end{align}
where the multiplet-averaged Auger yield $\overline{\xi}$ and
solid-angle correction factor $\overline{G}_\tau$, defined in
Eq.~(7), are
\begin{equation}
\overline{\xi} = \frac{\sum_J \hat{J}^2 \xi_J}{\hat{S}^2\hat{L}^2}, \quad \text{and} \quad \overline{G}_\tau = \frac{\sum_J \hat{J}^2 G_J \xi_J}{\sum_J \hat{J}^2 \xi_J}, \label{apx:eq_barGLxi}
\end{equation}

For the $^4\!P_J$ multiplet with $L=1$ and $S=3/2$, the statistical
weight denominator is
\mbox{$\hat{L}^2\hat{S}^2=(2L+1)(2S+1)=3\cdot4=12$}.
Using the weights $\hat{J}^2=\{2,4,6\}$ for
$J=\{1/2,3/2,5/2\}$, Eq.~\eqref{eq:hybrid_regrouped} becomes
\begin{equation}
\frac{d\sigma^{\text{hybrid}}_A[^4\!P]}{\overline{G}_\tau\,d\Omega_0}(0^\circ) = \overline{\xi}\,\frac{\sigma[^4\!P]}{4\pi} \Bigl( 1 + \alpha_2\, \mathcal{A}_{20}\,\overline{H}_2 \Bigr),
\label{eq:dealigned_hybrid_baseline}
\end{equation}
which is the hybrid form of the main-text master expression,
Eq.~(14), with
$M_2=\overline{H}_2(E_p)$. Here,
\begin{equation}
\overline{H}_2 \equiv \overline{H}_2(E_p) = \overline{H}_2[^4\!P] = \frac{2}{5}\left(\frac{\frac{\sigma[^4\!P]}{\hat{L}^2\,\hat{S}^2}\cdot \left(6 G_{5/2}\xi_{5/2} - 4 G_{3/2}\xi_{3/2}\right)}{ \overline{G}_\tau\cdot\overline{\xi}\cdot\sigma[^4\!P]}\right) = \frac{1}{15}\left(\frac{3G_{5/2}\xi_{5/2}-2 G_{3/2}\xi_{3/2}}{\overline{G}_\tau\cdot\overline{\xi}}\right).
\label{apx:eq_fraction_X_collapse}
\end{equation}
This is Eq.~(13) of the main text.
The solid-angle correction factors $G_J=G_J(E_p)$ are given as a
function of $E_p$ in
Tables~\ref{tab:cross_sections_final_13node_v23_python_clean} and
\ref{tab:computed_gj_comparison} and plotted in
Fig.~\ref{fig:GJGtau}. The hybrid projection factor
$\overline{H}_2(E_p)$ is plotted as a function of $E_p$ in
Fig.~1 and tabulated in
Tables~\ref{tab:model_a2_TOTAL_comparison} and \ref{tab:model_a2_SEC_only_comparison}.

\subsection{Mapping to Experimental Configurations with Isotropic Cascades}

To apply the pure $LS$ and hybrid $LS/LSJ$ treatments to the experimental
ZAPS configuration, the partial cross sections are explicitly separated
into direct SEC ($\sigma^{\text{SEC}}$) and indirect cascade
($\sigma^{\text{casc}}$) contributions. Using the isotropic-cascade
conditions established in Sec.~\ref{apx:sec_cs_defs}, the cross sections
become:
\begin{align}
\sigma(1,0) &\longrightarrow \sigma^{\text{tot}}(1,0)
= \sigma^{\text{SEC}}(1,0) + \sigma^{\text{casc}}(1,0),
\label{eq:sigma10}\\
\sigma(1,|1|) &\longrightarrow \sigma^{\text{tot}}(1,|1|)
= \sigma^{\text{SEC}}(1,|1|) + \sigma^{\text{casc}}(1,0),
\label{eq:sigma11}\\
\sigma^{\text{SEC}}[1]
&\longrightarrow \sigma^{\text{SEC}}(1,0)
+2\sigma^{\text{SEC}}(1,|1|),
\label{eq:sigmaSEC10plus11}\\
\sigma^{\text{casc}}[1]
&\longrightarrow \sigma^\text{casc}(1,0)
+2\sigma^\text{casc}(1,|1|)
=3\sigma^\text{casc}(1,0),
\label{eq:sigmacasc10plus11}\\
\sigma[1] &\longrightarrow \sigma^{\text{tot}}[1]
= \sigma^\text{SEC}[1]+\sigma^\text{casc}[1]
=\sigma^{\text{SEC}}[1]+3\sigma^\text{casc}(1,0).
\label{eq:sigmatotSECpluscasc}
\end{align}
For brevity, we omit the $[1]$ label hereafter and use
$\sigma\equiv\sigma[1]$, $\sigma^\text{tot}\equiv\sigma^\text{tot}[1]$,
$\sigma^\text{SEC}\equiv\sigma^\text{SEC}[1]$, and
$\sigma^\text{casc}\equiv\sigma^\text{casc}[1]$, so that
$\sigma^\text{tot}=\sigma^\text{SEC}+\sigma^\text{casc}$.

Using the $L=1$ pure-$LS$ alignment parameter
$\mathcal{A}_{20}$ from Eq.~\eqref{apx:eq_calA20Leq1_final_reduced} and
$\alpha_2=\alpha_2[L=1]=-\sqrt{2}$ from
Eq.~\eqref{apx:eq_alpha2_Leq1_val}, together with
Eqs.~\eqref{eq:sigma10}--\eqref{eq:sigmatotSECpluscasc}, gives
\begin{align}
\alpha_2\,\mathcal{A}_{20}^\text{tot}
&=\alpha_2\,\mathcal{A}_{20}^\text{tot}[L=1]
=(-\sqrt{2})(\sqrt{2})
\left[
\frac{\sigma^\text{tot}(1,|1|)
-\sigma^\text{tot}(1,0)}
{\sigma^\text{tot}[1]}
\right],
\label{eq:a2calA20_totpartials}\\
&=-2
\left[
\frac{\sigma^{\text{SEC}}(1,|1|)
-\sigma^{\text{SEC}}(1,0)}
{\sigma^\text{SEC}[1]}
\right]
\left(
\frac{\sigma^\text{SEC}[1]}
{\sigma^\text{tot}[1]}
\right),
\label{eq:a2calA20_SECpartialsprodratio}\\
&=\alpha_2\,\mathcal{A}^\text{SEC}_{20}
\left(
\frac{\sigma^\text{SEC}}
{\sigma^\text{tot}}
\right),
\label{eq:a2calA20_SECprodratio}
\end{align}
where
\begin{align}
\alpha_2\,\mathcal{A}^\text{SEC}_{20}
=-2\left(
\frac{\sigma^\text{SEC}(1,|1|)
-\sigma^\text{SEC}(1,0)}
{\sigma^\text{SEC}}
\right).
\label{apx:eq_alpha2_calA20_SEC}
\end{align}
Thus the isotropic cascade contribution reduces the total alignment
through the SEC fraction $\sigma^\text{SEC}/\sigma^\text{tot}$.

For the pure $LS$ treatment, corresponding to $M_2=1$ in
Eq.~(14), substitution of
Eq.~\eqref{eq:a2calA20_SECprodratio} gives
\begin{align}
\frac{d\sigma^{\text{LS tot}}_A[^4\!P]}
{\overline{G}_\tau\,d\Omega_0}(0^\circ)
&=\overline{\xi}\,
\frac{\sigma^\text{tot}[^4\!P]}{4\pi}
\left[
1+\alpha_2\,\mathcal{A}_{20}^\text{tot}
\right],
\nonumber\\
&=\overline{\xi}\,
\frac{\sigma^\text{tot}[^4\!P]}{4\pi}
\left[
1+\alpha_2\,\mathcal{A}^\text{SEC}_{20}
\left(
\frac{\sigma^\text{SEC}}
{\sigma^\text{tot}}
\right)
\right],
\nonumber\\
&=\frac{\overline{\xi}}{4\pi}
\left[
3\sigma^{\text{SEC}}(1,0)
+3\sigma^{\text{casc}}(1,0)
\right].
\label{eq:experimental_LS_final}
\end{align}
Thus the familiar $M_L=0$ filtering of the zero-degree pure-$LS$
distribution is retained for the directly populated SEC component,
whereas the isotropic cascade contribution remains unchanged.

Correspondingly, for the hybrid $LS/LSJ$ treatment,
$M_2=\overline{H}_2(E_p)$, and
\begin{align}
\frac{d\sigma^{\text{hybrid tot}}_A[^4\!P]}
{\overline{G}_\tau\,d\Omega_0}(0^\circ)
&=\overline{\xi}\,
\frac{\sigma^\text{tot}[^4\!P]}{4\pi}
\left[
1+\alpha_2\,\mathcal{A}_{20}^\text{tot}\,
\overline{H}_2
\right],
\nonumber\\
&=\overline{\xi}\,
\frac{\sigma^\text{tot}[^4\!P]}{4\pi}
\left[
1+\alpha_2\,\mathcal{A}^\text{SEC}_{20}
\left(
\frac{\sigma^\text{SEC}}
{\sigma^\text{tot}}
\right)
\overline{H}_2
\right],
\label{eq:SDCS0_hybrid_a2A20SEC_ratio}\\
&=\frac{\overline{\xi}}{4\pi}
\left[
(1+2\overline{H}_2)\sigma^{\text{SEC}}(1,0)
+2(1-\overline{H}_2)\sigma^{\text{SEC}}(1,|1|)
+3\sigma^{\text{casc}}(1,0)
\right].
\label{eq:experimental_hybrid_final}
\end{align}
The cascade term $3\sigma^{\text{casc}}(1,0)$ is isotropic and identical
in the pure-$LS$ and hybrid expressions. The difference between the two
treatments therefore arises entirely from the direct SEC contribution
through the hybrid projection factor $\overline{H}_2(E_p)$.

As shown in Sec.~\ref{apx:sub_eval_LSJ},
$\overline{H}_2(E_p)$ vanishes at
$E_p=6.26$~MeV. For $\overline{H}_2=0$, the hybrid AAD is isotropic and
the zero-degree prompt-equivalent Auger SDCS reduces to
$\overline{\xi}\,\sigma^\text{tot}/(4\pi)$, although the
collision-produced orbital alignment remains nonzero.

In the ideal unweighted limit, where $G_J\rightarrow1$ and
$\xi_J\rightarrow\overline{\xi}$, the hybrid projection factor becomes
\begin{equation}
H_2^\text{ideal}=\frac{1}{15}.
\label{eq:A_all_one}
\end{equation}
Equation~\eqref{eq:experimental_hybrid_final} then reduces to
\begin{equation}
\frac{d\sigma^\text{hybrid tot}_A}
{\overline{G}_\tau\,d\Omega_0}(0^\circ)
=
\frac{\overline{\xi}}{4\pi}
\left[
\frac{17}{15}\,\sigma^\text{SEC}(1,0)
+\frac{28}{15}\,\sigma^\text{SEC}(1,|1|)
+3\sigma^{\text{casc}}(1,0)
\right],
\label{apx:eq_final_hybrid_prompt_limit}
\end{equation}
which makes explicit the redistribution of the zero-degree SEC
contribution produced by the $LS/LSJ$ projection relative to the
pure-$LS$ limit.

The zero-degree prompt-equivalent SDCS obtained from
Eqs.~\eqref{eq:experimental_LS_final} and
\eqref{eq:experimental_hybrid_final} are plotted in
Fig.~2 for $M_2=1$ and
$M_2=\overline{H}_2(E_p)$, respectively. The ideal hybrid limit
$M_2=H_2^\text{ideal}=1/15$ is also shown for comparison.
The energy-dependent hybrid projection factor
$\overline{H}_2(E_p)$ is given by
Eq.~\eqref{apx:eq_fraction_X_collapse} and tabulated in
Tables~\ref{tab:model_a2_TOTAL_comparison} and \ref{tab:model_a2_SEC_only_comparison}.

\begin{table*}[tbh]
\caption{Calculated prompt-equivalent zero-degree Auger SDCS for the
metastable $1s2s2p\,^4\!P_J$ states in
$\text{C}^{4+}+\text{He}$ collisions, comparing the pure-$LS$ and
$LS/LSJ$ hybrid treatments for the total (SEC + cascades) contribution.
Also listed are the partial SEC cross sections
$\sigma^\text{SEC}_{10}\equiv\sigma^\text{SEC}(1,0)$ and
$\sigma^\text{SEC}_{1|1|}\equiv\sigma^\text{SEC}(1,|1|)$, the total
SEC cross section $\sigma^\text{SEC}$, the cascade cross section
$\sigma^\text{casc}_{10}\equiv\sigma^\text{casc}(1,0)$, and the total
cross section $\sigma^\text{tot}=\sigma^\text{SEC}+\sigma^\text{casc}$,
all from Ref.~\cite{mad22a}. The individual and multiplet-averaged
solid-angle correction factors $G_J$ and $\overline{G}_\tau$ are also
listed, together with
$\alpha_2\mathcal{A}_{20}^\text{SEC}$,
$\sigma^\text{SEC}/\sigma^\text{tot}$, and the ratio
$r_\text{hybrid}$ of the hybrid to pure-$LS$ zero-degree SDCS.
Relevant governing equations are indicated below the corresponding
columns.}
\label{tab:cross_sections_final_13node_v23_python_clean}
\centering
\begin{ruledtabular}
\begin{tabular}{c|ccccc|cc|cccc|cc|c}
$E_p$ & $\sigma^\text{SEC}_{10}$ & $\sigma^\text{SEC}_{1|1|}$ &
$\sigma^\text{SEC}$ & $\sigma^\text{casc}_{10}$ &
$\sigma^\text{tot}$ &
\(\alpha_2\mathcal{A}_{20}^\text{SEC}\) &
$\frac{\sigma^\text{SEC}}{\sigma^\text{tot}}$ &
$G_{1/2}$ & $G_{3/2}$ & $G_{5/2}$ & $\overline{G}_\tau$ &
$\frac{d\sigma_A^\text{LS tot}}{\overline{G}_\tau\,d\Omega_0}$ &
$\frac{d\sigma_A^\text{hyb tot}}{\overline{G}_\tau\,d\Omega_0}$ &
$r_{\text{hybrid}}$ \\[1mm]
(MeV) & \multicolumn{5}{c|}{($10^{-21}$~cm$^2$)} & & & & & & &
\multicolumn{2}{c}{($10^{-21}$~cm$^2$/sr)} & \\
&&&&&&\eqref{apx:eq_alpha2_calA20_SEC} &&
\multicolumn{3}{c}{\eqref{eq:Gtau_J}\footnotemark[1]} &
\eqref{eq:Gtau_avrg}&
\eqref{eq:experimental_LS_final}\footnotemark[2] &
\eqref{eq:SDCS0_hybrid_a2A20SEC_ratio}&\\
\hline
0.5  & 2.27E4 & 5.47E4 & 1.32E5 & 1.52E4 & 1.78E5 & -0.485 & 0.743 & 0.970 & 1.19 & 2.37 & 1.74 & 9.04E3 & 1.32E4 & 1.462 \\
1.0  & 4.01E4 & 3.13E4 & 1.03E5 & 1.89E4 & 1.60E5 & 0.171  & 0.645 & 1.00  & 1.33 & 2.30 & 1.76 & 1.41E4 & 1.29E4 & 0.917 \\
2.0  & 2.39E4 & 8.88E3 & 4.17E4 & 2.52E4 & 1.17E5 & 0.720  & 0.355 & 1.07  & 1.60 & 2.18 & 1.80 & 1.17E4 & 9.63E3 & 0.822 \\
3.0  & 1.28E4 & 3.15E3 & 1.91E4 & 8.03E3 & 4.32E4 & 1.01   & 0.442 & 1.13  & 1.85 & 2.06 & 1.84 & 4.97E3 & 3.58E3 & 0.719 \\
4.0  & 5.98E3 & 1.34E3 & 8.66E3 & 3.73E3 & 1.99E4 & 1.07   & 0.436 & 1.18  & 2.09 & 1.94 & 1.86 & 2.32E3 & 1.62E3 & 0.700 \\
5.0  & 2.93E3 & 642.   & 4.21E3 & 1.70E3 & 9.31E3 & 1.09   & 0.452 & 1.24  & 2.31 & 1.83 & 1.89 & 1.11E3 & 752.   & 0.680 \\
6.0  & 1.59E3 & 333.   & 2.26E3 & 795.   & 4.64E3 & 1.11   & 0.487 & 1.30  & 2.51 & 1.73 & 1.92 & 569.   & 371.   & 0.651 \\
7.0  & 919.   & 180.   & 1.28E3 & 413.   & 2.52E3 & 1.16   & 0.508 & 1.35  & 2.71 & 1.64 & 1.95 & 318.   & 198.   & 0.624 \\
8.0  & 548.   & 103.   & 754.   & 252.   & 1.51E3 & 1.18   & 0.499 & 1.41  & 2.88 & 1.55 & 1.97 & 191.   & 118.   & 0.615 \\
9.0  & 335.   & 63.9   & 463.   & 172.   & 979.   & 1.17   & 0.473 & 1.46  & 3.04 & 1.47 & 1.99 & 121.   & 75.5   & 0.623 \\
12.0 & 97.3   & 21.2   & 140.   & 54.9   & 305.   & 1.09   & 0.459 & 1.60  & 3.42 & 1.26 & 2.04 & 36.3   & 23.0   & 0.633 \\
15.0 & 32.8   & 8.28   & 49.4   & 18.4   & 105.   & 0.994  & 0.472 & 1.73  & 3.67 & 1.11 & 2.07 & 12.2   & 7.81   & 0.639 \\
18.0 & 12.4   & 3.76   & 19.9   & 7.03   & 41.0   & 0.867  & 0.486 & 1.85  & 3.77 & 1.02 & 2.08 & 4.64   & 3.07   & 0.661 \\
\end{tabular}
\footnotetext[1]{The $G_J$ solid-angle correction factors were
recalculated using Eq.~\eqref{eq:Gtau_J} and found to be in excellent
agreement with those used in Ref.~\cite{mad22a}.}
\footnotetext[2]{Traditional pure-$LS$ result, corresponding to $M_2=1$, as used
in Ref.~\cite{mad22a}.}
\end{ruledtabular}
\end{table*}

\begin{table}[htbp]
\caption{Calculated prompt-equivalent total zero-degree Auger SDCS
($10^{-21}$~cm$^2$/sr) for the metastable
$1s2s2p\,^4\!P_J$ states in $\text{C}^{4+}+\text{He}$ collisions.
Results are shown for the traditional pure-$LS$ treatment ($M_2=1$), the
Mehlhorn--Taulbjerg (MT) de-alignment treatment ($M_2=37/150$), the present
$LS/LSJ$ hybrid treatment
[$M_2=\overline{H}_2(E_p)$], and its ideal limit
($M_2=1/15$), together with the experimental data of
Ref.~\cite{mad22a}. The ratio of the pure-$LS$ to hybrid SDCS is also
listed. Results for $M_2=1$, $M_2=\overline{H}_2(E_p)$ and $M_2=1/15 \approx 0.0667$ are shown in Fig.~2.}
\label{tab:model_TOTAL_comparison_with_experiment}
\centering
\small
\setlength{\tabcolsep}{1.8pt}
\begin{ruledtabular}
\begin{tabular}{c|c|c|cc|c|c}
$E_p$ &
Pure $LS$ &
MT &
\multicolumn{2}{c|}{$LS/LSJ$ hybrid} &
SDCS Ratio & Experiment \\
(MeV) &
$M_2=1$ &
$M_2=\frac{37}{150}$ &
$M_2=\overline{H}_2(E_p)$ &
$M_2=\frac{1}{15}$ &
$(M_2=1)/(M_2=\overline{H}_2(E_p))$ &
(Madesis 2022) \\[1mm]
\hline
0.5  & 9.04E3 & 1.28E4 & 1.32E4 & 1.38E4 & 0.685 & \text{---} \\
1.0  & 1.41E4 & 1.30E4 & 1.29E4 & 1.28E4 & 1.09  & \text{---} \\
2.0  & 1.17E4 & 9.89E3 & 9.63E3 & 9.49E3 & 1.21  & \text{---} \\
3.0  & 4.97E3 & 3.81E3 & 3.58E3 & 3.54E3 & 1.39  & \text{---} \\
4.0  & 2.32E3 & 1.76E3 & 1.62E3 & 1.63E3 & 1.43  & \text{---} \\
5.0  & 1.11E3 & 829.   & 7.52E2 & 765.   & 1.48  & \text{---} \\
6.0  & 569.   & 417.   & 3.71E2 & 383.   & 1.53  & $302 \pm 114$ \\
7.0  & 318.   & 229.   & 1.98E2 & 208.   & 1.61  & \text{---} \\
8.0  & 191.   & 137.   & 1.18E2 & 125.   & 1.62  & \text{---} \\
9.0  & 121.   & 88.3   & 7.55E1 & 80.8   & 1.60  & $36.0 \pm 22.0$ \\
12.0 & 36.3   & 27.1   & 2.30E1 & 25.0   & 1.58  & $19.0 \pm 8.8$ \\
15.0 & 12.2   & 9.26   & 7.81   & 8.58   & 1.56  & $17.0 \pm 20.0$ \\
18.0 & 4.64   & 3.59   & 3.07   & 3.36   & 1.51  & \text{---} \\
\end{tabular}
\end{ruledtabular}
\end{table}

\subsection{Mehlhorn--Taulbjerg De-alignment Treatment}
\label{apx:subsub_Dk_parameter}

For completeness, we also consider the established Mehlhorn--Taulbjerg (MT)
de-alignment treatment~\cite{mehl80a}. This treatment describes the
evolution of the initially produced $LS$ alignment due to fine-structure
precession during the lifetime of the state, including the coherence
between overlapping fine-structure components. It is therefore distinct
from the isolated-fine-structure projection underlying the present
$LS/LSJ$ hybrid framework and is not part of the main-text formulation.
We include it here to establish this distinction explicitly and to provide
the corresponding isolated-level limiting value $D_2=37/150$ for comparison
with the pure-$LS$ and hybrid results.

Following sudden collision-induced population at $t=0$, the different
fine-structure components evolve with frequencies
$\omega_{JJ'}=(E_J-E_{J'})/\hbar$. Integration over the exponential
decay of the state introduces the factor
$1/(1+\epsilon_{JJ'}^2)$, where
$\epsilon_{JJ'}\equiv\Delta E_{JJ'}/\Gamma$. The resulting
configuration-wide de-alignment parameter is
\begin{equation}
D_k =
\sum_{J,J'}
\frac{\hat{J}^2\,\hat{J'}^2}{\hat{S}^2}
\frac{
\begin{Bmatrix}
J & J' & k \\
L & L & S
\end{Bmatrix}^2
}{
1+\epsilon_{JJ'}^2
}.
\label{eq:Dk_def_full}
\end{equation}

In the completely overlapping limit,
$\Delta E_{JJ'}\ll\Gamma$, the fine-structure components remain
coherent and $D_k\rightarrow1$, recovering the pure-$LS$ limit.
Conversely, for isolated fine-structure levels,
$\Delta E_{\rm FS}\gg\Gamma$, the off-diagonal terms with $J\ne J'$
are strongly suppressed. With $\epsilon_{JJ}=0$, Eq.~\eqref{eq:Dk_def_full}
then reduces for the $1s2s2p\,^4\!P_J$ manifold
($L=1$, $S=3/2$) to
\begin{equation}
D_k =
\sum_{J=1/2,3/2,5/2}
\frac{(2J+1)^2}{2S+1}
\begin{Bmatrix}
J & J & k \\
1 & 1 & 3/2
\end{Bmatrix}^2.
\label{eq:Dk_def}
\end{equation}

For $k=2$, the individual fine-structure contributions give
\begin{align}
D_2
&=
\frac{16}{4}(0)^2
+\frac{256}{4}
\left(-\frac{\sqrt{6}}{15}\right)^2
+\frac{1296}{4}
\left(-\frac{\sqrt{14}}{30}\right)^2
\nonumber\\
&=
0+\frac{8}{75}+\frac{7}{50}
=
\frac{37}{150}
\approx0.2467.
\label{apx:eq_D2_eval}
\end{align}

For $L=1$, the triangular condition associated with the
$\{L\,L\,k\}=\{1\,1\,k\}$ coupling in the Wigner $6j$ symbol restricts
the allowed tensor ranks to $k\leq2$. Consequently,
$D_4=D_6=\cdots=0$, and $D_2$ is the only nonzero anisotropic
de-alignment parameter for the $^4\!P$ manifold.

The corresponding MT-modified $LS$ AAD is
\begin{equation}
\frac{d\sigma_A^{LS+\mathrm{MT}}}
{\overline{G}_\tau\,d\Omega_0}(\theta)
=
\overline{\xi}\,
\frac{\sigma[L]}{4\pi}
\Bigl[
1+\alpha_2\,\mathcal{A}_{20}\,D_2
P_2(\cos\theta)
\Bigr],
\label{eq:cross_section_mehlhorn_global}
\end{equation}
where $\mathcal{A}_{20}$ and $\alpha_2$ are the pure-$LS$ alignment
and intrinsic anisotropy parameters evaluated in
Sec.~\ref{apx:sub_eval_LS}. Thus $D_2$ modifies the anisotropic
component of the pure-$LS$ distribution while leaving the isotropic
component unchanged.

At zero degrees, using
$\mathcal{A}_{20}[L=1]$ from
Eq.~\eqref{apx:eq_calA20Leq1_final_reduced} and
$\alpha_2=-\sqrt{2}$ from
Eq.~\eqref{apx:eq_alpha2_Leq1_val}, the distribution becomes
\begin{equation}
\frac{d\sigma_A^{LS+\mathrm{MT}}[^4\!P]}
{\overline{G}_\tau\,d\Omega_0}(0^\circ)
=
\overline{\xi}\,
\frac{\sigma[^4\!P]}{4\pi}
\Biggl[
1-2D_2
\left(
\frac{\sigma(1,|1|)-\sigma(1,0)}{\sigma}
\right)
\Biggr].
\end{equation}
Using
$\sigma=\sigma(1,0)+2\sigma(1,|1|)$ gives
\begin{equation}
\frac{d\sigma_A^{LS+\mathrm{MT}}[^4\!P]}
{\overline{G}_\tau\,d\Omega_0}(0^\circ)
=
\frac{\overline{\xi}}{4\pi}
\Bigl[
(1+2D_2)\sigma(1,0)
+2(1-D_2)\sigma(1,|1|)
\Bigr].
\label{eq:dealigned_LS_final}
\end{equation}

For $D_2=1$, Eq.~\eqref{eq:dealigned_LS_final} reduces to the
pure-$LS$ result derived in Sec.~\ref{apx:sub_eval_LS}, for which only
the $M_L=0$ partial cross section contributes at $\theta=0^\circ$.
For the MT isolated-level limit $D_2=37/150$, both magnetic-substate
components contribute to the zero-degree AAD.

\begin{figure}[htbp]
\centering
\includegraphics[width=\linewidth]{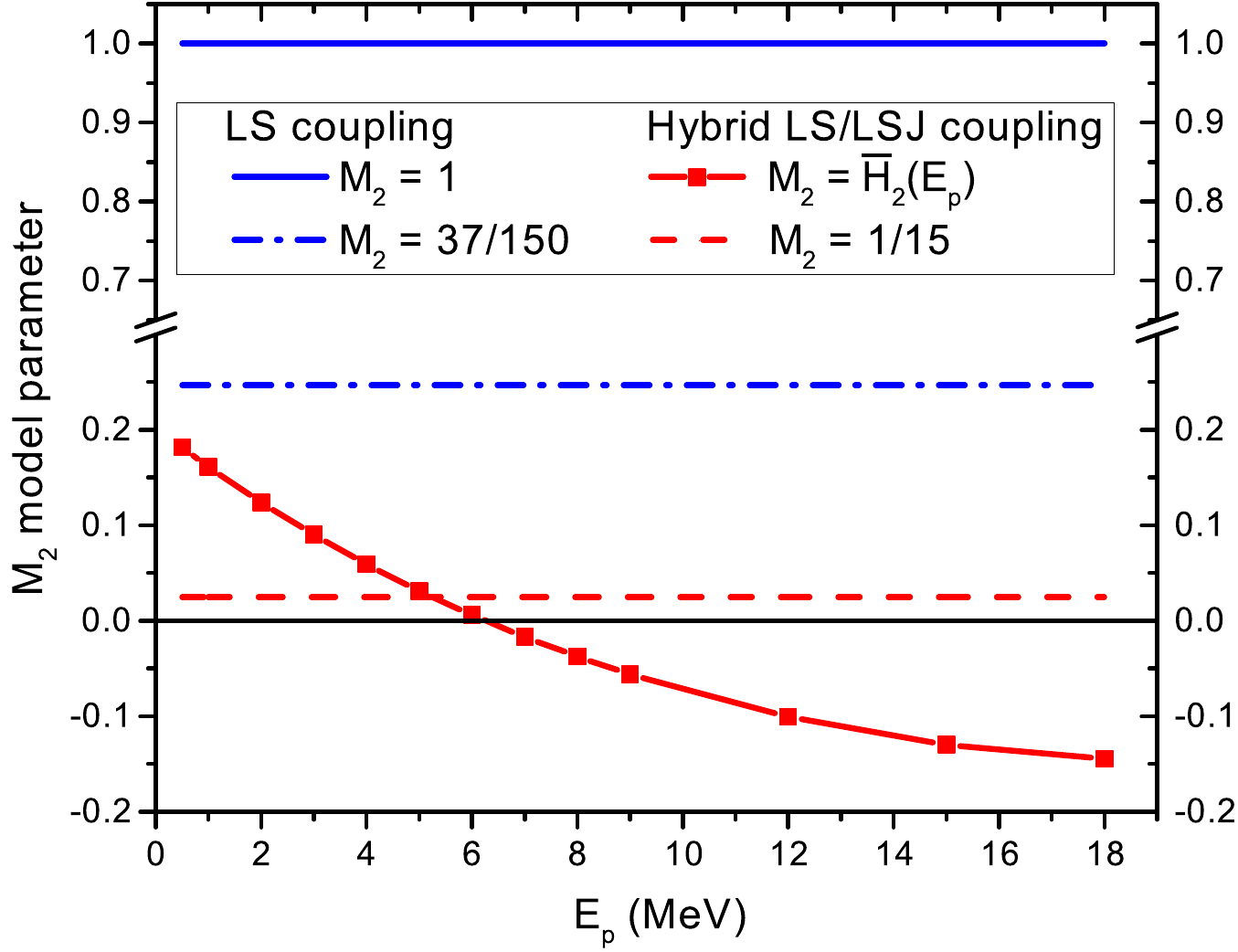}
\caption{$M_2$ vs projectile energy $E_p$ for the unresolved
$\text{C}^{3+}(1s2s2p\ ^4\!P)$ multiplet. The hybrid
$\overline{H}_2(E_p)$ crosses the isotropic line $M_2=0$ at
$E_p=6.26$~MeV. Also shown are the traditional pure-$LS$ result ($M_2=1$),
the Mehlhorn--Taulbjerg (MT) de-alignment treatment
($M_2=D_2=37/150 \approx 0.2467$), and the ideal
hybrid limit ($M_2=H_2^\text{ideal}=1/15 \approx 0.0667$).}
\label{fig:alignment_comparison_with_MT}
\end{figure}

Including the isotropic cascade contribution as established in the
preceding section, the total (SEC + cascades) zero-degree AAD becomes
\begin{equation}
\frac{d\sigma_A^{LS+\mathrm{MT},\,\text{tot}}[^4\!P]}
{\overline{G}_\tau\,d\Omega_0}(0^\circ)
=
\frac{\overline{\xi}}{4\pi}
\left[
\frac{224}{150}\,\sigma^{\text{SEC}}(1,0)
+\frac{226}{150}\,\sigma^{\text{SEC}}(1,|1|)
+3\sigma^{\text{casc}}(1,0)
\right].
\label{eq:experimental_LS_final_D2_eval}
\end{equation}
This is the MT result corresponding to
$M_2=D_2=37/150$ in Table~\ref{tab:model_TOTAL_comparison_with_experiment}.
Its relation to the pure-$LS$ result ($M_2=1$) and the present $LS/LSJ$
hybrid result [$M_2=\overline{H}_2(E_p)$] is illustrated in
Fig.~\ref{fig:alignment_comparison_with_MT}. The corresponding SEC-only
zero-degree Auger SDCS is shown in Fig.~\ref{fig:carbon_4P_SEC_only}.

\subsection{The fine-structure $H_k[L,S,J]$ hybrid coupling factor}
\label{apx:sub_HkLSJ}

The individual $J$ fine-structure hybrid coupling factor
$H_k[L,S,J]$ introduced in the main text [see Eq.~(11)] can be obtained explicitly
by combining the projection of the collision-produced $LS$ alignment
onto a given $J$ level [Eq.~(1)] with the
corresponding $LSJ$ intrinsic anisotropy parameter. For an individual
fine-structure level, the angular distribution parameter is
\begin{equation}
a_k[J] \equiv \alpha_{k}[J,J_f=0] \cdot
\mathcal{A}_{k0}^{\rm proj}[L,S,J].
\label{eq:akJ_hybrid_def}
\end{equation}

For autoionization transitions proceeding via single partial-wave
emission to a closed-shell final core ($J_f=0$, $L_f=0$), angular
momentum conservation gives $j=J$. As shown explicitly in
Sec.~\ref{apx:sec_deshalit_reduction}, application of the
de-Shalit--Talmi reduction formula to the general $LSJ$ intrinsic
anisotropy parameter gives
\begin{equation}
\alpha_{k} [J,0] =
(-1)^{J-\frac{1}{2}+k} \, \hat{J}\,
\hat{k}
\begin{pmatrix}
J & J & k \\
\frac{1}{2} & -\frac{1}{2} & 0
\end{pmatrix}.
\label{eq:alpha_j_collapsed}
\end{equation}
The corresponding pure-$LS$ intrinsic anisotropy parameter is
\begin{equation}
\alpha_{k} [L,0] =
(-1)^L  \, \hat{L}\,
\hat{k}
\begin{pmatrix}
L & L & k \\
0 & 0 & 0
\end{pmatrix}.
\label{eq:alpha_L_collapsed}
\end{equation}

The individual $J$ hybrid coupling factor $H_k[L,S,J]$ is related to
the pure-$LS$ quantities through
\begin{equation}
a_k[J] \equiv
\alpha_k[L,0]\,\mathcal{A}_{k0}\,H_k[L,S,J].
\end{equation}
Using the projected alignment
$\mathcal{A}_{k0}^{\rm proj}[L,S,J]$ from the main-text
Eq.~(1), together with
Eqs.~\eqref{eq:akJ_hybrid_def}, \eqref{eq:alpha_j_collapsed} and
\eqref{eq:alpha_L_collapsed}, gives
\begin{align}
H_k[L,S,J]
&=
\left[
(-1)^{L+S+J+k}\,\hat{L}\,\hat{J}
\begin{Bmatrix}
L & J & S \\
J & L & k
\end{Bmatrix}
\right]
\left[
\frac{\alpha_k[J,0]}{\alpha_k[L,0]}
\right]
\nonumber\\
&=
\left[
(-1)^{L+S+J+k}\,\hat{L}\,\hat{J}
\begin{Bmatrix}
L & J & S \\
J & L & k
\end{Bmatrix}
\right]
\left[
\frac{
(-1)^{J-\frac{1}{2}+k}\,\hat{J}\,
\hat{k}
\begin{pmatrix}
J & J & k \\
\frac{1}{2} & -\frac{1}{2} & 0
\end{pmatrix}
}{
(-1)^L\,\hat{L}\,
\hat{k}
\begin{pmatrix}
L & L & k \\
0 & 0 & 0
\end{pmatrix}
}
\right]
\nonumber
\intertext{Noting that $J$ is half-integer for the transitions considered
here, $2J-1$ is an even integer, we obtain}
H_k[L,S,J]&=
(-1)^{S+\frac{1}{2}}\,
\hat{J}^{2}
\begin{Bmatrix}
L & J & S \\
J & L & k
\end{Bmatrix}
\left[
\frac{
\begin{pmatrix}
J & J & k \\
\frac{1}{2} & -\frac{1}{2} & 0
\end{pmatrix}
}{
\begin{pmatrix}
L & L & k \\
0 & 0 & 0
\end{pmatrix}
}
\right].
\label{eq:Hk_master_uncontracted}
\end{align}

For the quadrupole rank $k=2$ of the
$1s2s2p\,^4\!P_J$ manifold ($L=1$, $S=3/2$),
Eq.~\eqref{eq:Hk_master_uncontracted} gives
\begin{align}
H_2\left[1, \frac{3}{2}, \frac{1}{2}\right]
&=
(2)\,
\begin{Bmatrix}
1 & 1/2 & 3/2 \\
1/2 & 1 & 2
\end{Bmatrix}
\left[
\frac{
\begin{pmatrix}
1/2 & 1/2 & 2 \\
1/2 & -1/2 & 0
\end{pmatrix}
}{
\begin{pmatrix}
1 & 1 & 2 \\
0 & 0 & 0
\end{pmatrix}
}
\right]
=0,
\label{eq:H2_12}
\\
H_2\left[1, \frac{3}{2}, \frac{3}{2}\right]
&=
(4)\,
\begin{Bmatrix}
1 & 3/2 & 3/2 \\
3/2 & 1 & 2
\end{Bmatrix}
\left[
\frac{
\begin{pmatrix}
3/2 & 3/2 & 2 \\
1/2 & -1/2 & 0
\end{pmatrix}
}{
\begin{pmatrix}
1 & 1 & 2 \\
0 & 0 & 0
\end{pmatrix}
}
\right]
=-\frac{2}{5},
\label{eq:H2_32}
\\
H_2\left[1, \frac{3}{2}, \frac{5}{2}\right]
&=
(6)\,
\begin{Bmatrix}
1 & 5/2 & 3/2 \\
5/2 & 1 & 2
\end{Bmatrix}
\left[
\frac{
\begin{pmatrix}
5/2 & 5/2 & 2 \\
1/2 & -1/2 & 0
\end{pmatrix}
}{
\begin{pmatrix}
1 & 1 & 2 \\
0 & 0 & 0
\end{pmatrix}
}
\right]
=+\frac{2}{5}.
\label{eq:H2_52}
\end{align}

Thus, for $J=1/2$, $3/2$, and $5/2$, the individual hybrid coupling factors are $H_2=0$, $-2/5$, and $+2/5$, respectively. The opposite signs of the $J=3/2$ and $J=5/2$
contributions are the origin of the suppression and possible inversion
of the unresolved anisotropy in the hybrid treatment. Their weighted combination gives the energy-dependent
$\overline{H}_2(E_p)$ defined in the main-text
Eq.~(13) and evaluated explicitly in
Eq.~\eqref{apx:eq_fraction_X_collapse}.

\subsection{Final results}

\begin{figure}[htbp]
\centering
\includegraphics[width=\columnwidth]{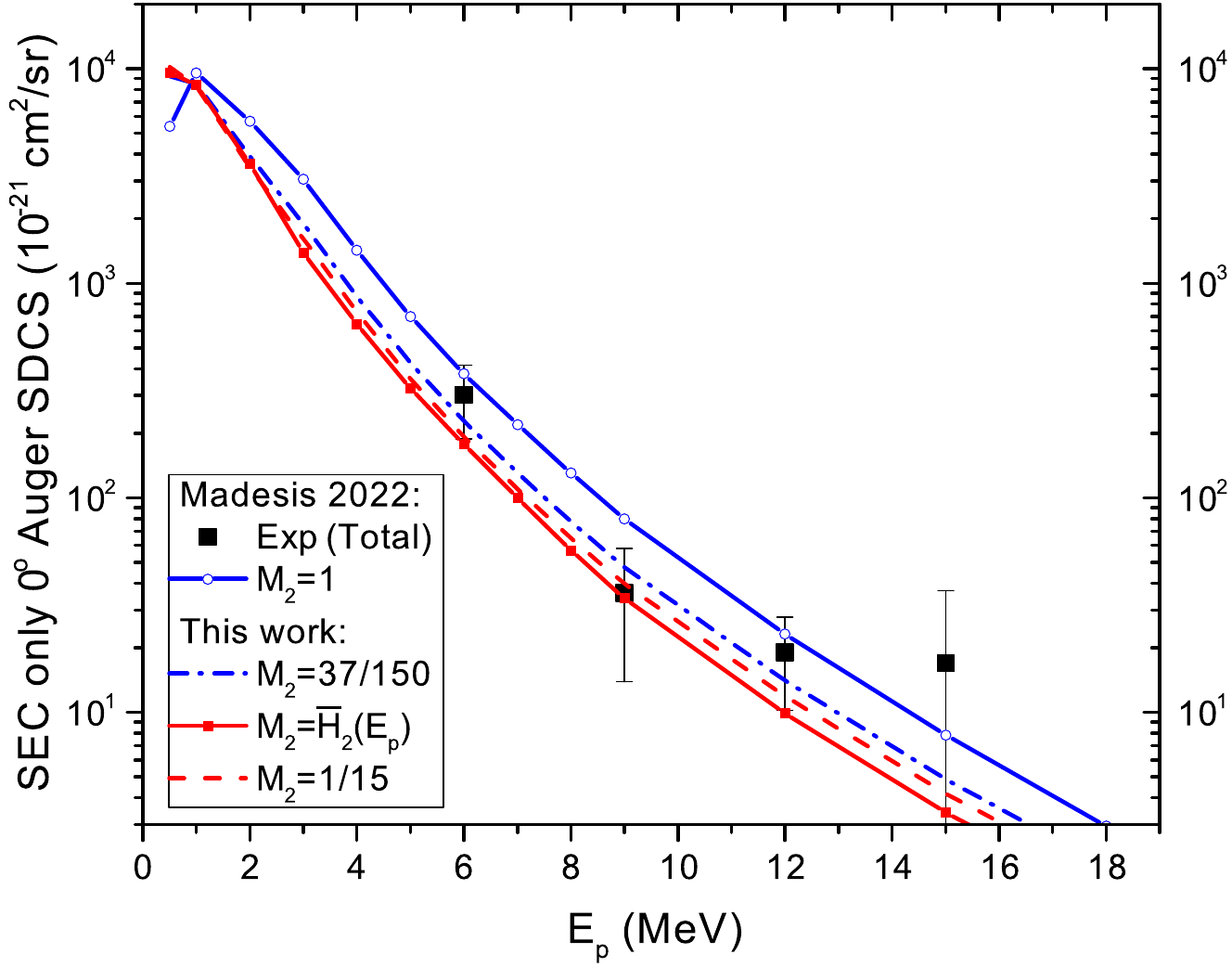}
\caption{Calculated SEC-only (no cascades) zero-degree Auger SDCS for the traditional
pure-$LS$ treatment ($M_2=1$), the Mehlhorn--Taulbjerg (MT) de-alignment treatment
($M_2=37/150$), and the hybrid $LS/LSJ$ treatment
($M_2=\overline{H}_2(E_p)$), together with the ideal hybrid limit ($M_2=1/15$).
The experimental points, which contain both SEC and cascade contributions, are shown
for comparison.}
\label{fig:carbon_4P_SEC_only}
\end{figure}

The corresponding SEC-only cross sections and the SEC-only and total
(SEC + cascades) $a_2$ parameters are summarized below. The SEC-only
cross sections are shown in Fig.~\ref{fig:carbon_4P_SEC_only} and
Table~\ref{tab:model_SEC_only_comparison}, while the $a_2$ parameters
are tabulated in Tables~\ref{tab:model_a2_TOTAL_comparison} and
\ref{tab:model_a2_SEC_only_comparison}.

\begin{table}[htbp]
\caption{Same as Table~\ref{tab:model_TOTAL_comparison_with_experiment},
but for SEC only (no cascades), including the cross-section ratio of the
pure-$LS$ treatment ($M_2=1$) to the present hybrid treatment
($M_2=\overline{H}_2(E_p)$). Results are also shown in
Fig.~\ref{fig:carbon_4P_SEC_only}.}
\label{tab:model_SEC_only_comparison}
\centering
\small
\setlength{\tabcolsep}{2.2pt}
\begin{ruledtabular}
\begin{tabular}{c|c|c|cc|c}
$E_p$ &
Pure $LS$ &
MT &
\multicolumn{2}{c|}{$LS/LSJ$ hybrid} &
SDCS Ratio \\
(MeV) &
$M_2=1$\footnotemark[1] &
$M_2=\frac{37}{150}$ &
$M_2=\overline{H}_2(E_p)$ &
$M_2=\frac{1}{15}$ &
$(M_2=1)/(M_2=\overline{H}_2(E_p))$ \\[1mm]
\hline
0.5  & 5.41E3 & 9.25E3 & 9.60E3 & 1.02E4 & 0.563 \\
1.0  & 9.57E3 & 8.52E3 & 8.41E3 & 8.27E3 & 1.14  \\
2.0  & 5.70E3 & 3.91E3 & 3.60E3 & 3.47E3 & 1.58  \\
3.0  & 3.06E3 & 1.90E3 & 1.39E3 & 1.62E3 & 2.20  \\
4.0  & 1.43E3 & 872.   & 6.47E2 & 738.   & 2.21  \\
5.0  & 700.   & 426.   & 3.24E2 & 360.   & 2.16  \\
6.0  & 379.   & 229.   & 1.78E2 & 193.   & 2.13  \\
7.0  & 219.   & 131.   & 1.00E2 & 110.   & 2.19  \\
8.0  & 131.   & 77.5   & 5.70E1 & 64.7   & 2.30  \\
9.0  & 80.0   & 47.5   & 3.41E1 & 39.7   & 2.35  \\
12.0 & 23.2   & 14.1   & 9.89   & 11.9   & 2.35  \\
15.0 & 7.83   & 4.89   & 3.41   & 4.19   & 2.30  \\
18.0 & 2.96   & 1.92   & 1.39   & 1.68   & 2.13  \\
\end{tabular}
\footnotetext[1]{SEC partial cross sections used for this calculation
are from Ref.~\cite{mad22a}.}
\end{ruledtabular}
\end{table}

\begin{table}[htbp]
\caption{Total (SEC + cascades) angular distribution parameters
$a_2=\alpha_2\,\mathcal{A}_{20}^\text{tot}\,M_2$, together with the
weighted hybrid coupling factor
$\overline{H}_2(E_p)$ and the ratio of the pure-$LS$ to the present
hybrid $a_2$ values. Values are also plotted in
Fig.~3(b).}
\label{tab:model_a2_TOTAL_comparison}
\centering
\small
\setlength{\tabcolsep}{4.0pt}
\begin{ruledtabular}
\begin{tabular}{c|c|c|c|cc|c}
$E_p$ &
$\overline{H}_2(E_p)$ &
\multicolumn{4}{c|}{$a_2=\alpha_2\,\mathcal{A}_{20}^\text{tot}\,M_2$} &
$a_2$ Ratio \\
(MeV) & &
Pure $LS$ &
MT &
\multicolumn{2}{c|}{$LS/LSJ$ hybrid} & \\
\cline{3-6}
& &
$M_2=1$ &
$M_2=D_2=\frac{37}{150}$ &
$M_2=\overline{H}_2(E_p)$ &
$M_2=H_2^\text{ideal}=\frac{1}{15}$ &
$(M_2=1)/(M_2=\overline{H}_2(E_p))$ \\ [1mm]
\hline
0.5   &  0.1815 & -0.361 & -0.0890 & -0.0656  & -0.0240  & 5.50  \\
1.0   &  0.1612 &  0.110 &  0.0271 &  0.0177  &  0.00734 & 6.21  \\
2.0   &  0.1240 &  0.256 &  0.0631 &  0.0318  &  0.0171  & 8.05  \\
3.0   &  0.0905 &  0.447 &  0.110  &  0.0405  &  0.0298  & 11.0  \\
4.0   &  0.0590 &  0.468 &  0.115  &  0.0276  &  0.0312  & 17.0  \\
5.0   &  0.0308 &  0.492 &  0.121  &  0.0151  &  0.0328  & 32.6  \\
6.0   &  0.0060 &  0.541 &  0.133  &  0.0033  &  0.0360  & 164.  \\
7.0   & -0.0170 &  0.586 &  0.145  & -0.0100  &  0.0391  & -58.6 \\
8.0   & -0.0376 &  0.589 &  0.145  & -0.0221  &  0.0393  & -26.7 \\
9.0   & -0.0560 &  0.554 &  0.137  & -0.0310  &  0.0369  & -17.9 \\
12.0  & -0.1004 &  0.500 &  0.123  & -0.0502  &  0.0333  & -9.96 \\
15.0  & -0.1296 &  0.469 &  0.116  & -0.0608  &  0.0313  & -7.71 \\
18.0  & -0.1443 &  0.421 &  0.104  & -0.0607  &  0.0281  & -6.94 \\
\end{tabular}
\end{ruledtabular}
\end{table}

\begin{table}[htbp]
\caption{Same as Table~\ref{tab:model_a2_TOTAL_comparison}, but for SEC
only (no cascades), with
$a_2=\alpha_2\,\mathcal{A}_{20}^\text{SEC}\,M_2$. Values are also plotted
in Fig.~3(a).}
\label{tab:model_a2_SEC_only_comparison}
\centering
\small
\setlength{\tabcolsep}{4.0pt}
\begin{ruledtabular}
\begin{tabular}{c|c|c|c|cc|c}
$E_p$ &
$\overline{H}_2(E_p)$ &
\multicolumn{4}{c|}{$a_2=\alpha_2\,\mathcal{A}_{20}^\text{SEC}\,M_2$} &
$a_2$ Ratio \\
(MeV) & &
Pure $LS$ &
MT &
\multicolumn{2}{c|}{$LS/LSJ$ hybrid} & \\
\cline{3-6}
& &
$M_2=1$ &
$M_2=D_2=\frac{37}{150}$ &
$M_2=\overline{H}_2(E_p)$ &
$M_2=H_2^\text{ideal}=\frac{1}{15}$ &
$(M_2=1)/(M_2=\overline{H}_2(E_p))$ \\ [1mm]
\hline
0.5   &  0.1815 & -0.485 & -0.120  & -0.0880  & -0.0323 & 5.51  \\
1.0   &  0.1612 &  0.171 &  0.0422 &  0.0276  &  0.0114 & 6.20  \\
2.0   &  0.1240 &  0.720 &  0.178  &  0.0893  &  0.0480 & 8.06  \\
3.0   &  0.0905 &  1.01  &  0.249  &  0.0914  &  0.0673 & 11.1  \\
4.0   &  0.0590 &  1.07  &  0.264  &  0.0631  &  0.0715 & 17.0  \\
5.0   &  0.0308 &  1.09  &  0.269  &  0.0336  &  0.0725 & 32.4  \\
6.0   &  0.0060 &  1.11  &  0.274  &  0.0067  &  0.0741 & 166.  \\
7.0   & -0.0170 &  1.16  &  0.286  & -0.0197  &  0.0770 & -58.9 \\
8.0   & -0.0376 &  1.18  &  0.291  & -0.0444  &  0.0787 & -26.6 \\
9.0   & -0.0560 &  1.17  &  0.289  & -0.0655  &  0.0781 & -17.9 \\
12.0  & -0.1004 &  1.09  &  0.269  & -0.1094  &  0.0726 & -9.96 \\
15.0  & -0.1296 &  0.994 &  0.245  & -0.1288  &  0.0663 & -7.72 \\
18.0  & -0.1443 &  0.867 &  0.214  & -0.1251  &  0.0578 & -6.93 \\
\end{tabular}
\end{ruledtabular}
\end{table}

\FloatBarrier
\section{Comprehensive Atomic Structure Dataset}

Table~\ref{tb:1s2s2p4Pas} compiles the underlying atomic structure parameters computed using
the relativistic Breit-Pauli framework~\cite{ben15a}. The natural width of each
fine-structure level is determined by its total Auger and radiative decay
rates, $\Gamma_J=\hbar(A^a_J+\sum_f A^x_J[f])=\hbar/\tau_J$, and the
corresponding Auger yield is
$\xi_J=A^a_J/(A^a_J+\sum_f A^x_J[f])$.
The statistically averaged natural width is
$\overline{\Gamma}=(2\Gamma_{1/2}+4\Gamma_{3/2}+6\Gamma_{5/2})/12$.
For the compact $J$-resolved presentation in Table~S6, the fine-structure
energy separations $|\Delta E_{JJ'}|$ are compared with
$\overline{\Gamma}$, giving
$\epsilon_{JJ'}=|\Delta E_{JJ'}|/\overline{\Gamma}$.
These ratios differ from those in Table~I of the main text, where the
pair-averaged width $\Gamma_{JJ'}=(\Gamma_J+\Gamma_{J'})/2$ is used
for the IRA criterion. Both measures give $\epsilon_{JJ'}\gg1$,
demonstrating that the three $^4\!P_J$ fine-structure levels are well
isolated on the scale of their natural widths. The unresolved Auger
intensity can therefore be treated as an incoherent sum of the individual
$J$ contributions, as assumed in the hybrid $LS/LSJ$ treatment.

\begin{table}
\caption{\label{tb:1s2s2p4Pas}Atomic structure parameters for the
C$^{3+}(1s2s2p\,^4\!P_J)$ states with $S=3/2$, $L=1$, and
$J={1/2,3/2,5/2}$. $A^x_J[f]$ and $A^a_J$ are the radiative and Auger
transition rates, respectively, $\xi_J$ are the corresponding Auger
yields, and $\overline{\xi}$ is the statistically averaged Auger yield.
$D_2$ is the MT de-alignment factor~\cite{mehl80a}.}

\begin{center}
\begin{tabular}{lccc}
\hline\hline
           C$^{3+}(1s2s2p\,^4\!P_J)$:                   & $J=1/2$ & $J=3/2$ & $J=5/2$ \\
\hline
$E_\text{cg}$ (eV)           & \multicolumn{3}{c}{294.095084\footnotemark[1]} \\
$E_J$ (eV)                    & 294.088644\footnotemark[1] & 294.089188\footnotemark[1] & 294.101161\footnotemark[1] \\
$\Delta E_{1/2,J}$ (eV)       & 0. & 5.44228E-4\footnotemark[1] & 125.1724E-4\footnotemark[1] \\
$Q_J$ (eV)                    & 44.779742\footnotemark[2] & 44.779198\footnotemark[2] & 44.767225\footnotemark[2] \\
\hline
$\sum_f A^x_J[f]$ (s$^{-1}$) & 1.45E6\footnotemark[3] & 3.83E5\footnotemark[3] & 1.91E4\footnotemark[3] \\
$A^a_J$ (s$^{-1}$)            & 3.39E8\footnotemark[3] & 1.37E8\footnotemark[3] & 8.22E6\footnotemark[3] \\
\hline
$\xi_J$                       & 0.9957\footnotemark[3] & 0.9972\footnotemark[3] & 0.9977\footnotemark[3] \\
$\overline{\xi}$ [Eq.~\eqref{apx:eq_barGLxi}]
                              & \multicolumn{3}{c}{0.9972\footnotemark[3]} \\
\hline
$\tau_J$ (s)                  & 2.94E-9\footnotemark[3] & 7.10E-9\footnotemark[3] & 121.36E-9\footnotemark[3] \\
$\Gamma_J$ (eV)               & 224.E-9\footnotemark[3] & 92.7E-9\footnotemark[3] & 5.42E-9\footnotemark[3] \\
$\overline{\Gamma}$ (eV)      & \multicolumn{3}{c}{70.94E-9\footnotemark[3]} \\
\hline
$|\Delta E_{JJ^\prime}|$ (eV) & 5.44228E-4\footnotemark[4] & 1.25172E-2\footnotemark[5] & 1.19730E-2\footnotemark[6] \\
$\epsilon_{JJ^\prime}=|\Delta E_{JJ^\prime}|/\overline{\Gamma}\qquad$
& 7.672E03\footnotemark[4]
& 1.764E05\footnotemark[5]
& 1.688E05\footnotemark[6] \\
$D_2$ [Eq.~\eqref{apx:eq_D2_eval}]
                              & \multicolumn{3}{c}{$\frac{37}{150} \approx 0.2467$\footnotemark[3]} \\
\hline\hline
\end{tabular}
\end{center}

\footnotetext[1]{\cite{yer17a,*yer17b} The center-of-gravity energy
$E_\text{cg}=21.61559(8)$~Ry is with respect to the $1s^22s$ ground
state. Using $1~\mathrm{Ry}=13.605693123$~eV gives
$E_\text{cg}=294.095084$~eV. The individual level energies are obtained
from $E_\text{cg}=[2E_{1/2}+4E_{3/2}+6E_{5/2}]/12$, with
$E_{3/2}-E_{1/2}=0.00004$~Ry and
$E_{5/2}-E_{3/2}=0.00088$~Ry.}

\footnotetext[2]{The SEC energy defect is calculated as
$Q_J=I_p(\mathrm{C}^{3+})+
E_\mathrm{exc}[\mathrm{C}^{4+}(1s2s\,^3\!S_1)]
-E_J-I_p(\mathrm{He})$, using
$I_p(\mathrm{C}^{3+})=4.74019(10)$~Ry~\cite{yer17a,*yer17b},
$E_\mathrm{exc}[\mathrm{C}^{4+}(1s2s\,^3\!S_1)]=298.962205$~eV,
and $I_p(\mathrm{He})=24.587389$~eV from the NIST Atomic Spectra
Database~\cite{kra26a}, where $I_p$ denotes the ionization potential.
The conversion $1~\mathrm{Ry}=13.605693123$~eV is used.}

\footnotetext[3]{\cite{ben15a}}
\footnotetext[4]{$J,J^\prime=1/2,3/2$}
\footnotetext[5]{$J,J^\prime=1/2,5/2$}
\footnotetext[6]{$J,J^\prime=3/2,5/2$}

\end{table}

\FloatBarrier
\section{The de-Shalit--Talmi Reduction Formula}
\label{apx:sec_deshalit_reduction}

The Wigner $3j$--$6j$ reduction identity given in de-Shalit and
Talmi~\cite[Appendix, p.~519]{des63a} is used to eliminate the
orbital-angular-momentum dependence in the single-partial-wave
expression for the $LSJ$ intrinsic anisotropy parameter
$\alpha^{(j)}_k[J,J_f]$. It leads to the reduced total-$j$ expression
used in the main text and subsequently in the derivation of the
individual $J$ hybrid coupling factor $H_k[L,S,J]$ in
Sec.~\ref{apx:sub_HkLSJ}. In the form required here, the identity is
\begin{align}
\hat{l}_1 \hat{l}_2 \begin{pmatrix}
          l_1 & l_2 & k \\
          0 & 0 & 0
        \end{pmatrix} \begin{Bmatrix}
          j_1 & j_2 & k \\
          l_2 & l_1 & \frac12
        \end{Bmatrix} & = - \begin{pmatrix}j_1 & j_2 & k \\
          \frac12 & -\frac12 & 0
        \end{pmatrix}\,\delta_{k, \, \text{even}},\label{apx:eq_deshalit}
\end{align}
where $\hat{X}\equiv\sqrt{2X+1}$ and the single-particle orbital angular momenta satisfy
$l_i=j_i\pm\frac{1}{2}$ through the triangular conditions
$\Delta(l_1j_1\frac{1}{2})$ and $\Delta(l_2j_2\frac{1}{2})$ of the
$6j$ symbol. The zero-projection $3j$ symbol requires
$l_1+l_2+k$ to be even. For parity-conserving transitions,
$l_1+l_2$ is even, so that only even $k$ contribute. Since this
restriction is not explicit on the right-hand side of
Eq.~\eqref{apx:eq_deshalit}, it is represented by
$\delta_{k,\,\mathrm{even}}$.
For the single-partial-wave emission limit of the $LSJ$ intrinsic
anisotropy parameter $\alpha_k[J,J_f]$~\cite[Eq.~(8)]{che92a},
\begin{align}
\alpha^{(l,j=(J))}_{k}[J, J_f=(0)] &= (-1)^{J+(0)-\frac{1}{2}} \, \hat{J} \, \hat{k} \, \hat{l}^2 \, \hat{(J)}^2
\begin{pmatrix} l & l & k \\ 0 & 0 & 0 \end{pmatrix}
\begin{Bmatrix} (J) & (J) & k \\ l & l & \frac{1}{2} \end{Bmatrix} \begin{Bmatrix} J & J & k \\ (J) & (J) & (0) \end{Bmatrix},
\label{eq:alpha_lJ_unreduced}
\end{align}
Setting $l=l'=l$ and $j=j'=j$ and applying
Eq.~\eqref{apx:eq_deshalit} to the $3j$--$6j$ product in brackets gives
\begin{align}
\alpha^{(l,j=(J))}_{k}[J, J_f=(0)] &= (-1)^{J+(0)-\frac{1}{2}} \, \hat{J} \, \hat{k} \, \hat{(J)}^2
\left[\hat{l}^2 \, \begin{pmatrix} l & l & k \\ 0 & 0 & 0 \end{pmatrix}
\begin{Bmatrix} (J) & (J) & k \\ l & l & \frac{1}{2} \end{Bmatrix}\right] \begin{Bmatrix} J & J & k \\ (J) & (J) & (0) \end{Bmatrix},\nonumber\\
&= (-1)^{J-\frac{1}{2}} \, \hat{J} \, \hat{k} \, \hat{(J)}^2
\left[- \begin{pmatrix}J & J & k \\
          \frac12 & -\frac12 & 0
        \end{pmatrix}\right] \begin{Bmatrix} J & J & k \\ (J) & (J) & (0) \end{Bmatrix},\nonumber\\
&= (-1)^{J+\frac{1}{2}} \, \hat{J} \, \hat{k} \, \hat{(J)}^2
\begin{pmatrix}J & J & k \\\frac12 & -\frac12 & 0\end{pmatrix}
\begin{Bmatrix} J & J & k \\ (J) & (J) & (0)\end{Bmatrix}.
\label{eq:alpha_lJ_reduced}
\end{align}
Using the identity for a $6j$ symbol with one zero,
\begin{align}
\begin{Bmatrix}
        j_1 & j_2 & 0 \\
       j_4 & j_5 & j_6
      \end{Bmatrix}=  \begin{Bmatrix}
        j_1 & j_5 & j_6 \\
       j_4 & j_2 &  0
      \end{Bmatrix}= &\, \frac{(-1)^{j_1+j_4+j_6}}{\hat{j_1}\,\hat{j_4}}\,\delta_{j_1j_2}\delta_{j_4j_5}.\label{eq:6-jonezero}
\end{align}
Defining
$\alpha^{(J)}_k[J]\equiv
\alpha^{(l,j=(J))}_k[J,J_f=(0)]$, we obtain
\begin{align}
\alpha^{(J)}_{k}[J]
= (-1)^{J+\frac{1}{2}} \, \hat{J} \, \hat{k} \, \hat{(J)}^2
\begin{pmatrix}J & J & k \\\frac12 & -\frac12 & 0\end{pmatrix}
\frac{(-1)^{2J+k}}{\hat{J}^2} = (-1)^{3J+k+\frac{1}{2}} \, \hat{J} \, \hat{k}
\begin{pmatrix}J & J & k \\\frac12 & -\frac12 & 0\end{pmatrix} = (-1)^{J-\frac{1}{2}+k} \, \hat{J} \, \hat{k}
\begin{pmatrix}J & J & k \\\frac12 & -\frac12 & 0\end{pmatrix}.
\label{eq:alpha_lJ_reduced_final}
\end{align}
This is the reduced single-partial-wave expression for the $LSJ$
intrinsic anisotropy parameter used in Eq.~(D1).

\section{Transformation of Partial Cross Sections from $LS$ to $LSJ$ Representations}
\label{apx:sec_LStoLSJ_cs_transformations}

For a spin-independent collision process, the population of magnetic
substates can be described either in the uncoupled $LS$ representation
or the coupled $LSJ$ fine-structure representation. The electron spin
$S$ acts as a spectator, with the spin substates $M_S$ populated
isotropically with equal statistical weight $1/\hat{S}^2$, where
$\hat{X}=\sqrt{2X+1}$.

To determine the fine-structure partial cross sections $\sigma(J,M_J)$
from a known set of orbital partial cross sections $\sigma(L,M_L)$,
the uncoupled states are projected onto the coupled basis.
Summing over the unpolarized spin projections gives the
Percival--Seaton transformation~\cite[Eq.~(3$\cdot$10)]{per58a} in
terms of a squared Wigner $3j$ symbol:
\begin{equation}
\sigma(J, M_J) = \frac{\hat{J}^2}{\hat{S}^2}
\sum_{M_L=-L}^{L}\sum_{M_S=-S}^{S}
\begin{pmatrix}
S & L & J \\
M_S & M_L & -M_J
\end{pmatrix}^2
\sigma(L, |M_L|).
\label{apx:eq_sigmaJMJ_to_sigmaLML}
\end{equation}
where conservation of angular momentum along the quantization axis
fixes the spin projection to $M_S=M_J-M_L$.

Summing over the total projection $M_J$ yields the total fine-structure
cross section $\sigma[J]$:
\begin{equation}
\sigma[J] = \sum_{M_J}\sigma(J,M_J)
= \frac{\hat{J}^2}{\hat{S}^2}
\sum_{M_J}\sum_{M_L}\sum_{M_S}
\begin{pmatrix}
S & L & J \\
M_S & M_L & -M_J
\end{pmatrix}^2
\sigma(L,|M_L|).
\end{equation}
Rearranging the order of summation gives
\begin{equation}
\sigma[J] = \frac{\hat{J}^2}{\hat{S}^2}
\sum_{M_L}
\left(
\sum_{M_J,M_S}
\begin{pmatrix}
S & L & J \\
M_S & M_L & -M_J
\end{pmatrix}^2
\right)
\sigma(L,|M_L|).
\end{equation}
Applying the standard Wigner $3j$ orthogonality relation to the
expression in parentheses~\cite[Eq.~(2.32)]{zar26a} gives
$1/\hat{L}^2$, leaving a direct relation to the total orbital production
cross section $\sigma[L]$:
\begin{equation}
\sigma[J] =
\frac{\hat{J}^2}{\hat{S}^2\,\hat{L}^2}
\sum_{M_L}\sigma(L,|M_L|)
=
\frac{\hat{J}^2}{\hat{S}^2\hat{L}^2}\,\sigma[L].
\label{apx:eq_sigmaJ_to_sigmaL_proof}
\end{equation}
This is the statistical-weight relation used in
Eq.~(8).

\subsection{Equivalence to the Percival--Seaton transformation}
\label{apx:sec_PS_projection_equivalence}

The Percival--Seaton transformation in
Eq.~\eqref{apx:eq_sigmaJMJ_to_sigmaLML}, when expressed in terms of
normalized alignment tensors, yields the geometrical $LS$-to-$LSJ$
projection used in Eq.~\eqref{apx:eq_calAk0_proj_def} and in the central
projection relation, Eq.~(1), of the main text.
The Percival--Seaton transformation contains more information than this
normalized projection alone. In addition to determining the relative
magnetic-sublevel populations, and hence the alignment, it also determines
the total population of each fine-structure level through the statistical-weight
relation derived above. Thus, the two results obtained from the
Percival--Seaton transformation provide separately the two quantities
$\sigma[J]$ and $\mathcal{A}_{k0}[J]$ required for the $J$-resolved AAD.

To show this, we define the alignment
parameter of a fine-structure level $J$ in terms of its magnetic-sublevel
cross sections~\cite[see Eq.~(5)]{ber77a} as
\begin{equation}
\mathcal{A}_{k0}[J] =
\frac{\hat{J}\,\hat{k}}{\sigma[J]}
\sum_{M_J=-J}^{J}
(-1)^{J-M_J}
\begin{pmatrix}
J & J & k \\
M_J & -M_J & 0
\end{pmatrix}
\sigma(J,|M_J|).
\label{apx:eq_calAk0J_PS}
\end{equation}
Substitution of Eq.~\eqref{apx:eq_sigmaJMJ_to_sigmaLML} gives
\begin{equation}
\mathcal{A}_{k0}[J] =
\frac{\hat{J}\,\hat{k}}{\sigma[J]}
\frac{\hat{J}^{2}}{\hat{S}^{2}}
\sum_{M_L,M_S,M_J}
(-1)^{J-M_J}
\begin{pmatrix}
J & J & k \\
M_J & -M_J & 0
\end{pmatrix}
\begin{pmatrix}
S & L & J \\
M_S & M_L & -M_J
\end{pmatrix}^{2}
\sigma(L,|M_L|).
\label{apx:eq_calAk0J_PS_sub}
\end{equation}

To evaluate the magnetic-sublevel sum, we use the contraction identity
of Ref.~\cite{zar26a}, Eq.~(2.60),
\begin{equation}
\begin{Bmatrix}
j_1 & j_2 & j_3 \\
j_4 & j_5 & j_6
\end{Bmatrix}
\begin{pmatrix}
j_5 & j_1 & j_6 \\
m_5 & m_1 & m_6
\end{pmatrix}
=
\sum_{m_2,m_3,m_4}
(-1)^{j_1+j_2-j_3+j_4+j_5+j_6-m_1-m_4}
\begin{pmatrix}
j_1 & j_2 & j_3 \\
m_1 & m_2 & -m_3
\end{pmatrix}
\begin{pmatrix}
j_4 & j_5 & j_3 \\
m_4 & m_5 & m_3
\end{pmatrix}
\begin{pmatrix}
j_2 & j_4 & j_6 \\
m_2 & m_4 & -m_6
\end{pmatrix}.
\label{apx:eq_Zare_contraction}
\end{equation}
For the present problem the required mapping is
\begin{equation}
(j_1,j_2,j_3,j_4,j_5,j_6)=(L,J,S,J,L,k), \qquad
(m_1,m_2,m_3,m_4,m_5,m_6)=(-M_L,M_J,M_S,-M_J,M_L,0).
\label{apx:eq_Zare_mapping}
\end{equation}
Although Eq.~\eqref{apx:eq_Zare_contraction} formally contains three
magnetic-quantum-number sums, they are not independent. The third $3j$
symbol requires $m_4=-m_2$, while either of the first two requires
$m_3=m_2-M_L$. Thus, for fixed $M_L$, $m_4=-M_J$ and
$m_3=M_S=M_J-M_L$, as required by angular-momentum conservation.

With the mapping in Eq.~\eqref{apx:eq_Zare_mapping}, the first two $3j$
symbols on the right-hand side of Eq.~\eqref{apx:eq_Zare_contraction}
can be written, using their symmetry properties, as
\begin{equation}
\begin{pmatrix}
L & J & S \\
-M_L & M_J & -M_S
\end{pmatrix}
=
(-1)^{L+S+J}
\begin{pmatrix}
S & L & J \\
M_S & M_L & -M_J
\end{pmatrix},
\qquad
\begin{pmatrix}
J & L & S \\
-M_J & M_L & M_S
\end{pmatrix}
=
(-1)^{L+S+J}
\begin{pmatrix}
S & L & J \\
M_S & M_L & -M_J
\end{pmatrix}.
\end{equation}
Their phase factors therefore cancel in the product. With the mapping
in Eq.~\eqref{apx:eq_Zare_mapping}, the remaining phase in
Eq.~\eqref{apx:eq_Zare_contraction} is
\begin{equation}
(-1)^{2L+2J-S+k+M_L+M_J}.
\end{equation}
This can be put into the required form since
\begin{equation}
\left(2L+2J-S+k+M_L+M_J\right)
-\left[(J-M_J)+(L+S+J+k)+(L-M_L)\right]
=2(M_L+M_J-S).
\end{equation}
Because $L$ and $M_L$ are integers and $J$ has the same
integer or half-integer character as $S$, $M_L+M_J-S$ is an integer.
The two exponents therefore differ by an even integer, and hence
\begin{equation}
(-1)^{2L+2J-S+k+M_L+M_J}
=
(-1)^{J-M_J}
(-1)^{L+S+J+k}
(-1)^{L-M_L}.
\label{apx:eq_PS_phase}
\end{equation}

Under the mapping in Eq.~\eqref{apx:eq_Zare_mapping}, the formal sum
$\sum_{m_2,m_3,m_4}$ in Eq.~\eqref{apx:eq_Zare_contraction} becomes
$\sum_{M_J,M_S,-M_J}$. Since the third $3j$ symbol restricts
$m_4$ to the single value $-M_J$, the $m_4$ sum contains only one
nonzero term and may be eliminated. The remaining formal sum can
therefore be written as $\sum_{M_J,M_S}$; moreover, for fixed $M_L$
the other $3j$ symbols enforce $M_S=M_J-M_L$.
Equation~\eqref{apx:eq_Zare_contraction} consequently gives
\begin{equation}
\sum_{M_J,M_S}
(-1)^{J-M_J}
\begin{pmatrix}
J & J & k \\
M_J & -M_J & 0
\end{pmatrix}
\begin{pmatrix}
S & L & J \\
M_S & M_L & -M_J
\end{pmatrix}^{2}
=
(-1)^{L+S+J+k}
\begin{Bmatrix}
L & J & S \\
J & L & k
\end{Bmatrix}
(-1)^{L-M_L}
\begin{pmatrix}
L & L & k \\
M_L & -M_L & 0
\end{pmatrix}.
\label{apx:eq_PS_contraction}
\end{equation}

Substitution into Eq.~\eqref{apx:eq_calAk0J_PS_sub} yields
\begin{equation}
\mathcal{A}_{k0}[J] =
(-1)^{L+S+J+k}
\frac{\hat{J}\,\hat{k}}{\sigma[J]}
\frac{\hat{J}^{2}}{\hat{S}^{2}}
\begin{Bmatrix}
L & J & S \\
J & L & k
\end{Bmatrix}
\sum_{M_L}
(-1)^{L-M_L}
\begin{pmatrix}
L & L & k \\
M_L & -M_L & 0
\end{pmatrix}
\sigma(L,|M_L|).
\end{equation}
From the definition of the orbital alignment parameter,
\begin{equation}
\sum_{M_L}
(-1)^{L-M_L}
\begin{pmatrix}
L & L & k \\
M_L & -M_L & 0
\end{pmatrix}
\sigma(L,|M_L|)
=
\frac{\sigma[L]}{\hat{L}\,\hat{k}}\,
\mathcal{A}_{k0}[L].
\end{equation}
Finally, using the statistical-weight relation
Eq.~\eqref{apx:eq_sigmaJ_to_sigmaL_proof}, we obtain
\begin{equation}
\mathcal{A}_{k0}[J]
=
(-1)^{L+S+J+k}
\hat{L}\hat{J}
\begin{Bmatrix}
L & J & S \\
J & L & k
\end{Bmatrix}
\mathcal{A}_{k0}[L].
\label{apx:eq_PS_projection_equivalence}
\end{equation}
Thus, in terms of the normalized alignment tensors, the
Percival--Seaton transformation reduces exactly to the geometrical
$LS$-to-$LSJ$ projection used in the present hybrid treatment.
For $L=1$ and $S=3/2$,
Eq.~\eqref{apx:eq_PS_projection_equivalence} reduces directly to
Eq.~\eqref{apx:eq_calAk0_proj_def}. The same relation also follows from
the general transformation of statistical tensors for coupled angular
momenta given in Ref.~\cite{bal00b}, Eq.~(1.69).

\section{Computation of solid angle correction factors}
\label{apx:sec_GJ_corrections}

The $^{4}\!P_J$ states are characterized by long, $J$-dependent
lifetimes~\cite{dav89a}. Thus, the three $J=1/2,3/2,5/2$
fine-structure levels undergo Auger emission along the projectile path
from the gas cell through the spectrometer assembly. The resulting
effective electron solid angle depends on the level lifetime $\tau_J$,
the projectile velocity $V_p$, and the geometric acceptance at each
point of emission. The spectrometer acceptance, including the effects of the injection
lens, was modeled in previous work by Monte Carlo ray-tracing
simulations with the SIMION~\cite{simion} ion-optics package~\cite{dou15a,ben18a}.

The correction factor $G_J$ is defined as the ratio of the effective
solid angle for a long-lived $J$ level to that for a prompt state
decaying within the gas cell. The overall factor $\overline{G}_\tau$
is obtained by averaging over the fine-structure components:
\begin{align}
\overline{\Delta\Omega} & = \overline{G}_\tau\,\overline{\Delta\Omega_0}(s_0,r_0), \label{eq:dOavrg}\\
\overline{G}_\tau & = \frac{\sum_J \hat{J}^2 \xi_J G_J}{\sum_J \hat{J}^2 \xi_J}, \label{eq:Gtau_avrg}\\
G_J & = \frac{\overline{\Delta\Omega_J}}{\overline{\Delta\Omega_0}(s_0,r_0)}, \label{eq:Gtau_J}
\end{align}
where $\hat{J}^2\equiv2J+1$. Following Ref.~\cite{dou15a}, the
point-source solid angle for a source on the symmetry axis at a distance
$s$ from the spectrometer entrance aperture is
\begin{equation}
\Delta\Omega_0(s, r) = 2\pi\left(1-\frac{s}{\sqrt{r^2+s^2}}\right), \label{eq:dOmegas}
\end{equation}
where $r$ is the effective acceptance radius.

In Ref.~\cite{dou15a}, the spatial averaging over a gas cell of length
$L_c$ is performed using the coordinate $z^\prime$, measured from the
upstream entry window of the cell. Here we introduce the shifted
excitation coordinate $z^{\prime\prime}=z^\prime-L_c/2$, placing the
origin at the center of the gas cell. The prompt solid angle, obtained by averaging the point-source solid
angle over the target gas-cell length, is
\begin{equation}
\overline{\Delta\Omega_0}(s_0,r_0) = \frac{1}{L_c}\int_{-L_c/2}^{L_c/2} \Delta\Omega_0(s_0 - z^{\prime\prime}, r_0) \, dz^{\prime\prime}, \label{eq:dOmega0avrg}
\end{equation}
where $s_0$ is the distance from the center of the gas cell to the
spectrometer entrance aperture of radius $r_0$. For our specific
geometry, we compute
$\overline{\Delta\Omega_0}=1.515\times10^{-4}$~sr, using
$L_c=52.5$~mm, $s_0=289$~mm, and $r_0=2$~mm~\cite{dou15a,mad22a}.

For the long-lived states, the post-excitation integration must account
for the finite angular acceptance of the spectrometer near the entrance
aperture. For pre-retardation operation at $F=4$, SIMION ray-tracing
simulations give a maximum accepted half-angle
$\theta_{\text{max}}=2^\circ$~\cite{ben18a}, corresponding to the
critical distance $z_0=r_0/\tan\theta_{\text{max}}$.

For a state excited at position $z^{\prime\prime}$, the post-excitation
travel distance $z$ is integrated from the excitation point ($z=0$)
to the spectrometer entrance aperture
($z=s_0-z^{\prime\prime}$). The integration is divided at
$z=s_0-z^{\prime\prime}-z_0$, where the remaining distance to the
aperture equals the critical distance $z_0$. The resulting averaged
solid angle for each fine-structure level is
\begin{equation}
\overline{\Delta\Omega_J} = \frac{1}{L_c}\int_{-L_c/2}^{L_c/2} dz^{\prime\prime} \left[ \int_{0}^{s_0 - z^{\prime\prime} - z_0} I_1(z, z^{\prime\prime}) \, dz + \int_{s_0 - z^{\prime\prime} - z_0}^{s_0 - z^{\prime\prime}} I_2(z, z^{\prime\prime}) \, dz \right], \label{eq:DOmegaJ}
\end{equation}

Here, $s=s_0-z^{\prime\prime}-z$ is the remaining distance to the
spectrometer entrance aperture. For
$z<s_0-z^{\prime\prime}-z_0$, corresponding to $s>z_0$, the collection
geometry is determined by the physical aperture radius $r_0$. In the
near-aperture region, $z\ge s_0-z^{\prime\prime}-z_0$, corresponding
to $s\le z_0$, the finite angular acceptance is included through the
effective radius $r(s)=s\tan\theta_{\text{max}}$. The corresponding
integrands are
\begin{align}
I_1(z, z^{\prime\prime}) &= \frac{\exp\left(-\frac{z}{V_p\tau_J}\right)}{V_p\tau_J} \,\Delta\Omega_0(s,r_0)
= \frac{\exp\left(-\frac{z}{V_p\tau_J}\right)}{V_p\tau_J} \,\Delta\Omega_0(s_0-z^{\prime\prime}-z,r_0), \label{eq:I1_def}\\
I_2(z, z^{\prime\prime}) &= \frac{\exp\left(-\frac{z}{V_p\tau_J}\right)}{V_p\tau_J} \,\Delta\Omega_0(s,s\tan\theta_{\text{max}})
= \frac{\exp\left(-\frac{z}{V_p\tau_J}\right)}{V_p\tau_J} \,\Delta\Omega_0\left(s_0-z^{\prime\prime}-z,(s_0-z^{\prime\prime}-z)\frac{r_0}{z_0}\right). \label{eq:I2_def}
\end{align}

\begin{figure}[t]
\centering
\includegraphics[width=\linewidth]{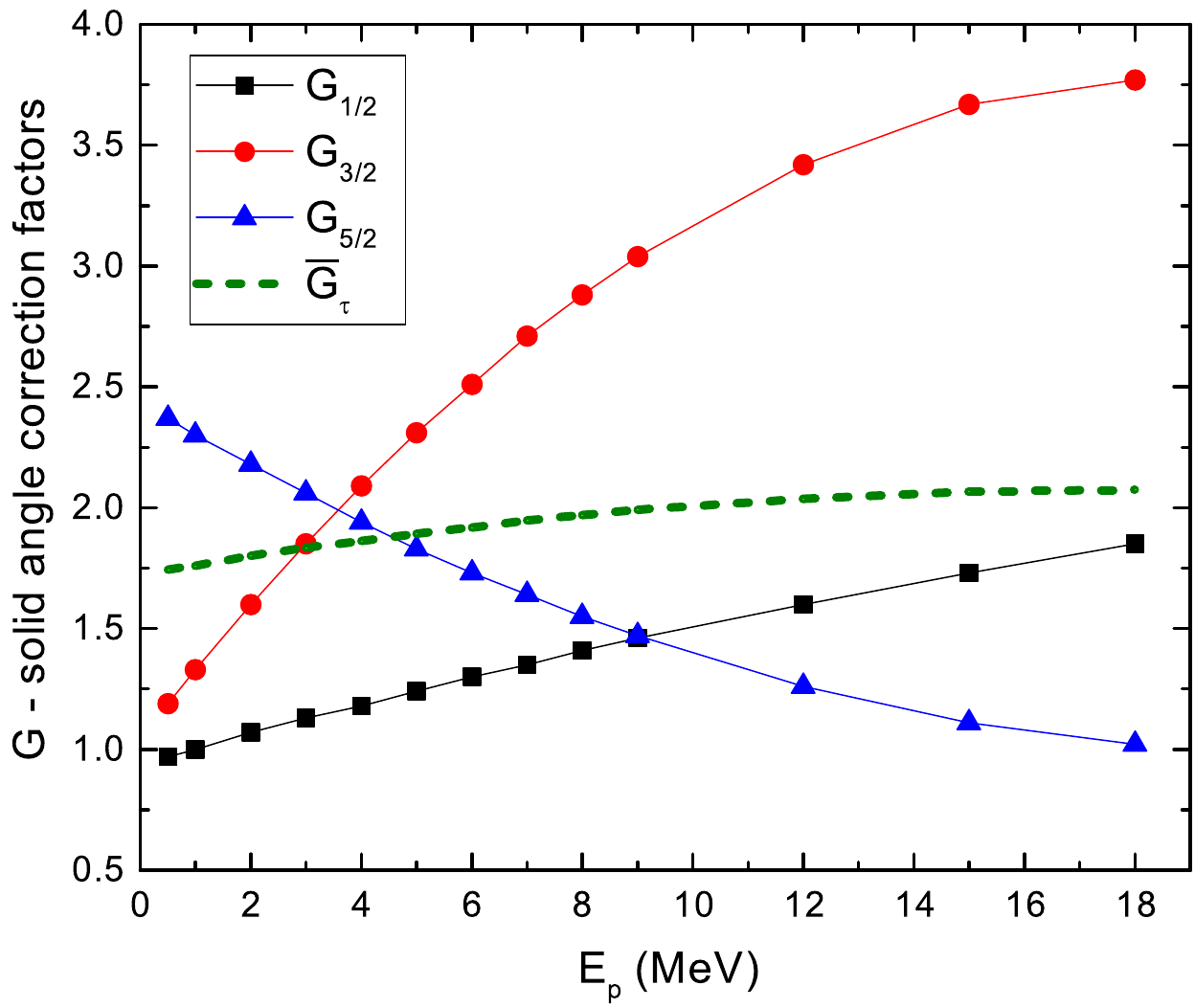}
\caption{\label{fig:GJGtau}
Solid-angle correction factors $G_J$ and their weighted average
$\overline{G}_\tau$ versus laboratory projectile energy $E_p$ for the
fine-structure components of C$^{3+}(1s2s2p\,^4\!P_J)$. Individual
$J$ components are shown by solid lines with markers, and
$\overline{G}_\tau$ by the dashed line. The solid-angle treatment
follows Refs.~\cite{dou15a,ben18a} and was used in the analysis of
Ref.~\cite{mad22a}. Numerical values are given in
Table~\ref{tab:computed_gj_comparison}, and the corresponding values
used in the prompt-equivalent SDCS are given in
Table~\ref{tab:cross_sections_final_13node_v23_python_clean}.}
\end{figure}

The evaluations use lifetimes of $\tau_{1/2}=2.94$~ns,
$\tau_{3/2}=7.10$~ns, and $\tau_{5/2}=121.36$~ns, along with the
respective Auger yields of $\xi_{1/2}=0.9957$, $\xi_{3/2}=0.9972$,
and $\xi_{5/2}=0.9977$ (Table~\ref{tb:1s2s2p4Pas}). The calculated
values of $V_p$, the characteristic decay lengths $V_p\tau_J$, the
individual factors $G_J$, and the averaged factor $\overline{G}_\tau$
are compiled in Table~\ref{tab:computed_gj_comparison} and plotted in
Fig.~\ref{fig:GJGtau}.

As shown in the figure, $G_{1/2}$ and $G_{3/2}$ increase with projectile
energy because the increasing decay lengths shift the spatial
distribution of Auger emission downstream toward regions of greater
effective solid-angle acceptance. In contrast, $G_{5/2}$ decreases with
increasing energy. Owing to the much longer lifetime
$\tau_{5/2}=121.36$~ns, its decay length greatly exceeds
$s_0=289$~mm over most of the energy range
(Table~\ref{tab:computed_gj_comparison}), so that an increasing fraction
of the Auger emission occurs beyond the region of favorable
spectrometer acceptance.

\begin{table}[h]
\caption{Calculated fine-structure solid-angle correction factors $G_J$
and their weighted average $\overline{G}_\tau$ for carbon ($M_p=12$~amu)
at projectile energies $E_p$. Also shown are the corresponding projectile
velocities $V_p$, the decay lengths $V_p\tau_J$, and the product
$\overline{G}_\tau\overline{\xi}$, where $\overline{G}_\tau$ and
$\overline{\xi}$ are defined by Eq.~(7), with
$\overline{\xi}=0.9972167$ obtained from Table~\ref{tb:1s2s2p4Pas}.
The solid-angle correction factors are plotted in
Fig.~\ref{fig:GJGtau}.}
\label{tab:computed_gj_comparison}
\centering
\setlength{\tabcolsep}{5.5pt}
\begin{tabular}{cc|ccc|ccc|cc}
\hline\hline
$E_p$ & $V_p$ & $V_p\tau_{1/2}$ & $V_p\tau_{3/2}$ & $V_p\tau_{5/2}$ &
\raisebox{-0.5ex}{$G_{1/2}$} & \raisebox{-0.5ex}{$G_{3/2}$} &
\raisebox{-0.5ex}{$G_{5/2}$} & \raisebox{-0.5ex}{$\overline{G}_\tau$} &
\raisebox{-0.5ex}{$\overline{G}_\tau\overline{\xi}$} \\
(MeV) & (mm/ns) & (mm) & (mm) & (mm) & & & & & \\
\hline
0.5  &  2.836 &   8.34 &  20.13 &  344.11 & 0.970 & 1.190 & 2.370 & 1.744 & 1.739 \\
1.0  &  4.010 &  11.79 &  28.47 &  486.63 & 1.000 & 1.330 & 2.300 & 1.760 & 1.755 \\
2.0  &  5.670 &  16.67 &  40.26 &  688.16 & 1.070 & 1.600 & 2.180 & 1.802 & 1.797 \\
3.0  &  6.944 &  20.42 &  49.30 &  842.76 & 1.130 & 1.850 & 2.060 & 1.835 & 1.830 \\
4.0  &  8.018 &  23.57 &  56.93 &  973.07 & 1.180 & 2.090 & 1.940 & 1.863 & 1.858 \\
5.0  &  8.964 &  26.35 &  63.64 & 1087.85 & 1.240 & 2.310 & 1.830 & 1.892 & 1.887 \\
6.0  &  9.819 &  28.87 &  69.71 & 1191.60 & 1.300 & 2.510 & 1.730 & 1.918 & 1.913 \\
7.0  & 10.605 &  31.18 &  75.29 & 1286.99 & 1.350 & 2.710 & 1.640 & 1.948 & 1.943 \\
8.0  & 11.336 &  33.33 &  80.49 & 1375.76 & 1.410 & 2.880 & 1.550 & 1.970 & 1.965 \\
9.0  & 12.023 &  35.35 &  85.36 & 1459.12 & 1.460 & 3.040 & 1.470 & 1.992 & 1.986 \\
12.0 & 13.880 &  40.81 &  98.55 & 1684.50 & 1.600 & 3.420 & 1.260 & 2.037 & 2.031 \\
15.0 & 15.515 &  45.62 & 110.16 & 1882.95 & 1.730 & 3.670 & 1.110 & 2.067 & 2.061 \\
18.0 & 16.993 &  49.96 & 120.65 & 2062.26 & 1.850 & 3.770 & 1.020 & 2.075 & 2.069 \\
\hline\hline
\end{tabular}
\end{table}

\end{document}